\documentclass[preprint,11pt,preprintnumbers,nofootinbib]{revtex4}
\usepackage{graphicx}% Include figure files
\usepackage{dcolumn}% Align table columns on decimal point
\usepackage{bm}% bold math
\usepackage{amsfonts}
\usepackage{amsmath,color,graphicx,mathrsfs,subfigure}

\def\circa#1{\,\raise.3ex\hbox{$#1$\kern-.75em\lower1ex\hbox{$\sim$}}\,}
\makeatletter

\usepackage{epsfig}
\usepackage{amsmath,amsfonts,amssymb}

\def\be{\begin{equation}}
\def\ee{\end{equation}}

\def\bea{\begin{eqnarray}}
\def\eea{\end{eqnarray}}

\def\zetai{\zeta_{\rm inf}}
\def\zetac{\zeta_{\rm curv}}
\def\zetat{\zeta_{\rm trap}} 

\def\fnl{f_{\rm NL}}
\def\nfnl{n_{\rm f_{NL}}}

\def\bk{\mathbf k}

\def\fnl{f_{\rm NL}}

\def\p{{\mbox{\tiny{ESP}}}}   %DL: new macro
\def\vp{\vec{\varphi}}
\def\beq{\begin{equation}}
\def\eeq{\end{equation}}
\begin{document}
%\hfill BI-TP/2011/15

\date{\today}
\title{Beauty is Distractive: \\
Particle production during multifield inflation}
\author{Diana Battefeld$^{1)}$} 
\email[email: ]{dbattefe(AT)astro.physik.uni-goettingen.de}
\author{Thorsten Battefeld$^{1)}$}
\email[email: ]{tbattefe(AT)astro.physik.uni-goettingen.de}
\author{Christian Byrnes$^{2)}$}
\email[email: ]{byrnes(AT)physik.uni-bielefeld.de}
\author{David Langlois$^{3)}$}
\email[email: ]{langlois(AT)apc.univ-paris7.fr}
\affiliation{1) Institute for Astrophysics,
University of Goettingen,
Friedrich Hund Platz 1,
D-37077 Gottingen, Germany}
\affiliation{2) Fakult{\"a}t f{\"u}r Physik, Universit{\"a}t Bielefeld,
Postfach 100131, 33501 Bielefeld, Germany}
\affiliation{3) APC (CNRS-Universit\'e Paris 7), \\
10, rue Alice Domon et L\'eonie Duquet, 75205 Paris Cedex 13, France}

\begin{abstract}
We consider a two-dimensional model of inflation, where the inflationary trajectory is ``deformed''  
by a grazing encounter with an Extra Species/Symmetry Point (ESP) {\it after} the observable cosmological scales have left the Hubble radius. The encounter entails a sudden production of particles, whose backreaction 
 causes a bending of the trajectory and a temporary decrease in speed, both of which are sensitive to initial conditions. This ``modulated" effect leads to an additional contribution to the curvature perturbation, which can be dominant   if the encounter is close. We compute associated non-Gaussianities, the bispectrum and its scale dependence as well as the trispectrum, which are potentially detectable in many cases. In addition, we consider a direct modulation of the coupling to the light field at the ESP via a modulaton field, a mixed scenario whereby the modulaton is identified with a second inflaton, and an extended Extra Species Locus (ESL); all of these scenarios lead to similar additional contributions to observables. We conclude that inflaton interactions throughout inflation are strongly constrained if primordial non-Gaussianities remain unobserved in current experiments such as PLANCK. If they are observed, an ESP encounter leaves additional signatures on smaller scales which may be used to identify the model.
\end{abstract}

\maketitle
\newpage

\tableofcontents

%%%%%%%%%%%%%%%%%%%%%%%%%%%%%%%%%%%%
%%%%%%%%%%%%%%%%%%%%%%%%%%%%%%%%%%%%
\section{Introduction}

In the simplest models of inflation, the inflaton smoothly rolls down its potential while continuously producing super-Hubble perturbations via amplification of quantum fluctuations. When trying to embed the inflaton into a more concrete framework based on high energy physics models, one finds that this uneventful 
story could be altered in several ways. 
It is important to identify these alternative possibilities and to relate them to specific observational signatures, that could be looked for in 
 future cosmological data. 

An example of such alterations is the possibility that the coupling of the inflaton to other fields might affect its evolution. This is in particular the case if some field coupled to the inflaton becomes suddenly light at some special point along  the inflationary trajectory. Indeed, whenever the effective mass of a field becomes small, there may be an associated burst of particle production, as thoroughly studied in preheating ~\cite{Kofman:1997yn} (see also~\cite{Dolgov:1989us,Traschen:1990sw}) where this occurs while the inflaton is oscillating at the bottom of its potential.

The same phenomenon can also take place during inflation, as pointed out in   
\cite{ckrt,Elgaroy:2003hp,Romano:2008rr}, and  such an event can  temporarily trap or at least slow-down the inflaton by draining its kinetic energy. If such a trapping event occurs while observable modes are exiting the horizon during inflation this may lead to features in the power-spectrum \cite{ckrt}, as well as  localised non-Gaussian spikes in the CMB, (see \cite{Barnaby:2010sq} for a recent comprehensive paper on this topic).
By contrast, if this event occurs well after all the observable modes have exited the Hubble radius  there might not be any observable signature (except perhaps primordial black hole formation on very small scales) if the trapping effect is exactly the same in all parts of our accessible universe. However, if the strength of the trapping fluctuates, this acts as an additional source of primordial adiabatic perturbations because the duration of inflation 
 varies from place to place \cite{Langlois:2009jp}. This modulation of the trapping can be the consequence of an explicit dependence of the coupling on another light scalar field as considered in \cite{Langlois:2009jp}. 

In the present work, we primarily focus on
the situation where the modulation is due to the fluctuations of the initial conditions of a multi-dimensional inflaton. Indeed, if one assumes that particle production occurs near a specific point in the multi-dimensional inflaton field space, then the trapping effect is
 stronger if the inflationary trajectory comes closer to this point during the grazing encounter. A slight change in the initial position of the inflaton  thus affects the whole duration of inflation, depending on how much particle production slows down the inflatons and/or  alters its initial trajectory. 

The location in field space where additional degrees of freedom become light are often associated with  enhanced symmetries,  which some authors consider beautiful \cite{Kofman:2004yc}. These extra species/symmetry locations (ESL, ESP if it is a point) are a common occurrence in moduli spaces originating in string theory, see  \cite{Seiberg:1994rs,Seiberg:1994aj,Witten:1995im,Intriligator:1995au,Strominger:1995cz,Witten:1995ex,Katz:1996ht,Bershadsky:1996nh,Witten:1995gx} for a small selection of examples, and moduli fields can  get trapped near such loci \cite{Kofman:2004yc,Watson:2004aq,Patil:2004zp,Patil:2005fi,Cremonini:2006sx,Greene:2007sa}  (the string Higgs effect \cite{Bagger:1997dv,Watson:2004aq}). If these fields drive inflation and if several ESPs/ESLs are encountered, trapped inflation can result  \cite{Kofman:2004yc,Green:2009ds,Silverstein:2008sg,Battefeld:2010sw} (see also \cite{Bueno Sanchez:2006eq,Bueno Sanchez:2006ah,Brax:2009hd,Matsuda:2010sm,Brax:2011si,Lee:2011fj} for related work); however, in this paper, we focus on a single grazing ESP encounter.
Since inflationary trajectories are slowed down and derail from their prior course as the ESP is passed, one might say that beauty is distractive in our case.
  
A modulated trapping event can easily produce significant non-Gaussianity of the local type, as stressed in  \cite{Langlois:2009jp}.
The current observational bound on
the local type of the bispectrum from seven years of WMAP data is $-10<\fnl<74$ at the $2\sigma$ level
\cite{Komatsu:2010fb}. The bounds on the trispectrum, which are roughly $\tau_{NL}\lesssim 10^5$ and $|g_{NL}|\lesssim 10^6$, come from large scale structure \cite{Desjacques:2009jb},
and the CMB \cite{Vielva:2009jz,Smidt:2010sv,Fergusson:2010gn}. If there is no detection, constraints with Planck should be tightened by an order of magnitude. Future CMB constraints on the trispectrum parameters were considered in \cite{Smidt:2010ra}.

We find that a single close ESP encounter during inflation, as discussed in Sec.~\ref{sec:grazingESP}, can yield the dominant contribution to the power-spectrum, see Sec.~\ref{sec:tunedESP}, if the coupling is of order one (we also discuss mixed scenarios, where $\Xi=\mathcal{P}_{new}/\mathcal{P}_{total}$ is neither close to zero nor one); this new contribution is accompanied by observably large non-Gaussianities. If the encounter occurs during the time-frame when large-scale modes are leaving the horizon, an additional bump-like feature in the power-spectrum results ruling out the model for $g\sim 1$ \cite{Barnaby:2009mc,Barnaby:2009dd,Barnaby:2010ke,Barnaby:2010sq}; however,  a subsequent encounter can successfully modulate large-scale fluctuations; the still present bump on smaller scales can, if observed in future experiments,
 be used to identify the time of the encounter and offer a consistency check. If non-Gaussianities are not observed, we can rule out the majority of ESP encounters, that is, we can severely constrain the interactions of the inflatons during inflation.

In models considered in this paper, the dominant contribution to observables originates from the slowing down of the inflatons, Sec.~\ref{sec:slowingdown}, whereas the bending of the trajectory can be neglected, Sec.~\ref{sec:geom}. We derive approximate analytic expressions for the power-spectrum and non-Gaussianities, which we compare with a full numerical solution in Sec.~\ref{sec:numerics}. We find that the analytic expressions are excellent for intermediate impact parameters $\mu\sim 10^{-3}M_{P}$, which are needed anyhow if the additional contribution to the power-spectrum satisfies the COBE bound. For smaller impact parameters temporary trapping events can take place if the coupling is strong enough; however, since trajectories become chaotic and thus extremely sensitive to initial conditions, we can rule out such encounters. A summary of the different regimes that we discuss is provided in Table~\ref{tab:regimes}. 

We relate the likelihood of  a single ESP encounter to the average inter ESP separation in (and the dimensionality of) field space in Sec.~\ref{Sec:trappedinflation}, where we also comment on the relationship to trapped inflation. Modest ESP densities with an average inter-ESP distance of $y\sim 0.2 M_{P}$ are sufficient in a two dimensional field space to render a grazing ESP encounter a likely occurrence during the last sixty efoldings of inflation. Thus, no particular tuning is required. 

A direct modulation of the coupling constant is considered in Sec.~\ref{sec:modulatedtrapping}. We compute the same observables in several cases, Sec.~\ref{sec:modtrapspecialcases}, and comment on the possibility to use the modulaton as a curvaton in Sec.~\ref{sec:modulatonascurvaton}. The two proposals, a grazing ESP encounter and modulated trapping, are closely related, but neither is a subset of the other. This is evident if the modulaton is promoted to a second inflaton, Sec.\ref{sec:grazESPandModulatedTrapping}. In all cases we find comparable non-Gaussianities, and we provide consistency relations between the various observables $n_s,r,f_{NL},n_{f_{NL}},\tau_{NL},g_{NL}$ throughout this article.

Some technical aspects related to the validity of the approximations used to describe particle production (non-adiabaticity, sudden event) at an ESP are provided in the appendix.

\section{A grazing ESP encounter \label{sec:grazingESP}}
We are interested in the effect that an ESP (Extra Species/Symmetry Point \cite{Kofman:2004yc,Watson:2004aq}) encounter has on the trajectory, and subsequently on observables, in a $2$D field space during inflation. An application of our results to higher dimensions is straightforward as long as only a single ESP encounter takes place. Throughout this section we set $M_{P} =1/\sqrt{(8\pi G)}=1$.

 \begin{figure*}[tb]
\begin{center}
\includegraphics[scale=0.5,angle=0]{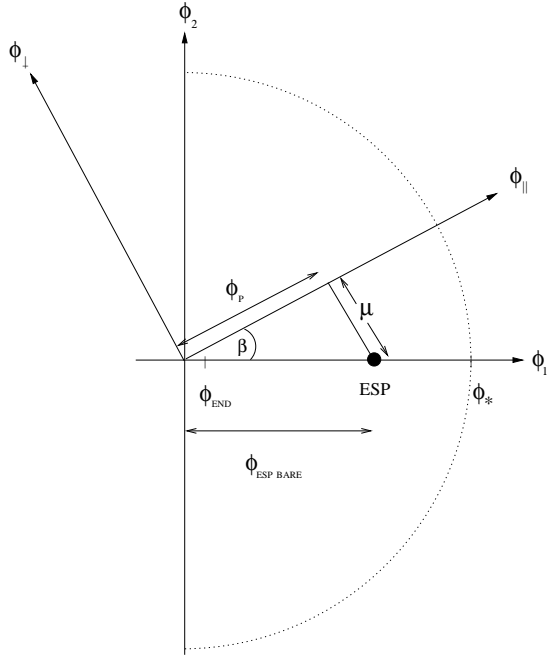}
   \caption{\label{pic:trajectory1} Schematic of a grazing Extra Species Point (ESP) encounter with impact parameter $\mu$ in a two dimensional field space. If the dimensionality is bigger than two, the depicted plane (and coordinates $\varphi_\perp,\varphi_\parallel$) is defined by  the location of the ESP and the tangent to the unperturbed trajectory at $\varphi_{\parallel}=\varphi_p$. Note that $\varphi_\perp$ is not an instantaneous isocurvature field, but defined via a global field redefinition. In case of a simple potential, as in (\ref{sqpot}), trajectories are straight lines in the absence of an ESP. $N=60$ efoldings of inflation take place between $|\vec{\varphi}|=\varphi_*$ and $|\vec{\varphi}|=\varphi_{\mbox{\tiny END}}$.   }
   \end{center}
\end{figure*}

\subsection{Particle production in a simple multi-field model}

\begin{figure*}[tb]
\begin{center}
\includegraphics[scale=0.5,angle=0]{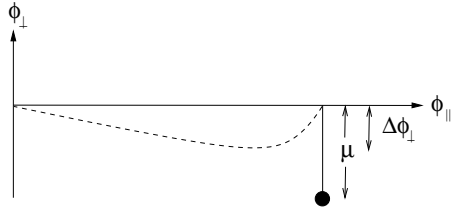}
   \caption{\label{pic:trajectory2} Schematic of the change in trajectory of a grazing ESP encounter (no trapping, see Fig.~\ref{pic:trajectories} (c)  for an example of a temporarily trapped trajectory): in the absence of an ESP the trajectory is aligned with the $\varphi_{\parallel}$ axis and the dotted line is the bent trajectory due to backreaction of $\chi$-particles onto the inflaton fields. We approximate the production of $\chi$-particles as a single, instantaneous event at the point of closest approach. The amount of bending (\ref{Deltavarphiperp2}) and slowing down (\ref{velocitydecrease}) depends sensitively on the impact parameter $\mu$, which impacts correlation functions. Actual trajectories are plotted in Fig.~\ref{pic:trajectories} (a).}
   \end{center}
\end{figure*}

Consider a simple multi-field inflationary model with  two inflatons $\varphi_i$, $i=1,2$, canonical kinetic terms and a quadratic potential
\begin{equation}
V(\varphi)=\frac{1}{2}m^2\sum_i\varphi_i^2\equiv \frac{1}{2}m^2\vec{\varphi}^2\label{sqpot}\,.
\end{equation} 
In addition, we assume a coupling to another scalar field $\chi$ via the Langrangian 
\begin{equation}
\mathcal L_{int}=-\frac{1}{2}g^2\chi^2(\vec{\varphi}-\vec{\varphi}_{\mbox{\tiny{ESP}}})^2\,.
\end{equation}
For $g>0.001$ radiative corrections to the effective potential are possible; these can be absent in super-symmetric theories where bosonic and fermionic contributions tend to cancel each other out \cite{Kofman:1997yn,Berera:1998cq}, or in models within string theory \cite{Kofman:2004yc,Green:2009ds,Silverstein:2008sg}. In the following we assume the absence of such corrections. 
When the inflationary trajectory comes close to the ESP, the $\chi$ particles become light
\begin{eqnarray}
m_{\chi}^{eff}=g|\vec{\varphi}-\vec{\varphi}_{\mbox{\tiny{ESP}}}|+m_{\chi}^{\mbox{\tiny bare}}
\end{eqnarray}
 (in the following we ignore the presence of a small bare mass for the $\chi$ particles). Thus, if the impact parameter 
\begin{eqnarray}
\mu=\mbox{min}(|\vec{\varphi}(t)-\vec{\varphi}_{\mbox{\tiny{ESP}}}|)
\end{eqnarray}
 is small enough (see Figures \ref{pic:trajectory1} and \ref{pic:trajectory2}), 
particle production can occur, similar to preheating \cite{Traschen:1990sw,Kofman:1997yn}, see \cite{Bassett:2005xm,Kofman:2008zz} for reviews and \cite{Battefeld:2008bu,Battefeld:2009xw,Braden:2010wd} for multi-field preheating. For small impact parameters this event is fast compared to the Hubble time during inflation, $\Delta t_p \ll H$, and can be treated as an instantaneous event in the following. 

The particle occupation number increase can be approximated by \cite{Kofman:1997yn,Kofman:2004yc} (see Appendix \ref{app:nonadd} for technical details)
\begin{equation}
n_p=\frac{g^{3/2}v^{3/2}_p}{(2\pi)^3}e^{-\frac{\pi g\mu^2}{v_p}}\, ,\label{n_p}
\end{equation}
where $v_p\equiv|\dot{\vec\varphi}(t_p)|$. The subscript $p$ denotes the time of particle production when the distance to the ESP is minimal.
Equation (\ref{n_p}) is an analytic approximation valid for $g>H_p^2/v_p$ \cite{Kofman:2004yc} (always satisfied in this paper).
 The number of particles is diluted by the subsequent expansion 
\begin{equation}
n\equiv n_p\left(\frac{a_p}{a}\right)^{3}\Theta(t-t_p)\,,
\end{equation}
 so that backreaction is confined to a few efoldings at most ($\Theta$ is the Heaviside distribution).

Particle production
is stronger the smaller the impact parameter $\mu$ is and the faster the trajectory is traversed. The velocity at which particle production occurs is easily found in the slow--roll regime when the slow-roll parameters $\varepsilon_i\equiv\left( {V_{,i}}/{ V}\right)^2/2$ and   $\eta_i\equiv{V_{,ii}}/{V}$ are small (we use the shorthand--notation $V_{,i}\equiv \partial V/\partial \varphi_i$).
Given the potential in (\ref{sqpot}) the total slow--roll parameters along the slow-roll
 trajectory are simply $\varepsilon_{\parallel}=\eta_{\parallel}={2}/{\vec{\varphi}^2}$  and we can solve the
approximate 
  Friedmann and Klein--Gordon equations to get the speed \begin{equation}
v_p\simeq \sqrt{\frac{2}{3}}m=\mbox{const} \,. \label{vp}
\end{equation}

In our scenario, what matters is not the particle production in itself, but the influence of the particle production onto the motion of the inflaton. We thus need to compute the backreaction  of these particles onto the inflatons. This is a complicated problem  in general, but, in our case, it is reasonable to  
 use the Hartree approximation, so that the equation of motion of the inflaton becomes
 (see Appendix (\ref{app:2}) for details) 
  \begin{equation}
\ddot\varphi_i+3H\dot\varphi_i+\frac{\partial V}{\partial\varphi_i}=-\frac{\partial\rho_{\chi}}{\partial\varphi_i}\,,\label{eom}
\end{equation}
where
\begin{equation}
\rho_{\chi}\approx n_p\left(\frac{a_p}{a}\right)^3m_{\chi}^{{eff}}\,,
\end{equation}
is the energy density of the $\chi$ field. This energy density originates from infusing kinetic energy of the inflatons into $\rho_{\chi}$, which is usually small\footnote{The decrease in kinetic energy is accounted for in all numerical studies in this paper.}, $\rho_\chi(t_p)\ll \sum_i\dot\varphi_i^2/2=v_p^2/2$.

\subsection{Nonlinear curvature perturbation}

As we will quantify in the following subsections, 
any change in the trajectory caused by the ESP encounter, be it a change in velocity or an actual deviation of the trajectory, leads to a change in the number of efoldings, $\Delta N$. This change is sensitive to the initial conditions around sixty efoldings before the end of inflation at $t_*$, which in turn causes additional contributions to correlation functions of the curvature perturbation on uniform density surfaces $\zeta$, most easily recovered in the (non-linear) $\delta N$-formalism  \cite{Lyth:2004gb,Lyth:2005fi};  these fluctuations are related to the fluctuations in the number of efoldings $\zeta=\delta N$, where
\begin{eqnarray}
N=\int_{t_*}^{t_{end}} H dt\,.
\end{eqnarray}
In our case, the total number of efoldings can be written as
\beq
N=N_{\mbox{\tiny{SR}}}+\Delta N,
\eeq
where $N_{\mbox{\tiny{SR}}}$ corresponds to the number of efoldings along the slow--roll trajectory in {\it absence} of an ESP, while $\Delta N$ represents the change in the number of efoldings due to the ESP encounter. 

In this section we provide the relevant expressions based on the $\delta N$-formalism for correlation functions, before computing $\Delta N$ numerically and, in certain regimes, analytically.

\subsubsection{The power-spectrum}
Since the masses  of $\varphi_i$  are smaller than the Hubble parameter, $m\ll H$, all $\varphi_i$ 
carry nearly scale invariant fluctuations on super Hubble scales with 
power-spectrum
\begin{equation}
\langle
\delta\varphi^i_{\bk_1} \delta\varphi^j_{\bk_2} \rangle = (2
\pi)^3 \delta_{ij} \delta^{(3)} (\bk_1 +
\bk_2) P(k_1), \qquad P(k)\equiv\frac{2 \pi^2}{k^3} \mathcal{P} (k), \quad 
\mathcal P=\left(\frac{H_*}{2\pi}\right)^2\,.\label{Pperp}
\end{equation}
(In this paper $\varphi_i=\varphi^i$). 
 The index $*$ denotes Hubble crossing of the pivot scale 
 $(k_{60}=a_*H_*)$ 
 at $N=60$ efoldings before the end of inflation. 
 
 Expanding the number of efoldings at first order as a function of the inflaton perturbations, 
 $\delta\varphi_i$, or alternatively $\delta\varphi_\parallel$ and $\delta\varphi_\perp$ (where $\varphi_\parallel$ and $\varphi_\perp$ are the components respectively parallel and orthogonal to the trajectory without particle production; see Fig.~\ref{pic:trajectory1}),
 the curvature perturbation can be written as
\begin{equation}
\zeta=\delta N=\delta N_{\mbox{\tiny{SR}}}+\delta\Delta N=N_{,\parallel} \delta\varphi_\parallel+ \Delta N_{,\perp}\delta\varphi_\perp\,,
\end{equation}
with the notation
\begin{eqnarray}
N_{,\parallel}\equiv \frac{\partial N}{\partial\varphi_\parallel^*} ,\qquad  \Delta N_{,\perp}\equiv\frac{\partial \Delta N}{\partial \varphi_{\perp}^*} \,,
\end{eqnarray}
(we employ corresponding expressions for  higher--order derivatives). 
Here we used that due to the geometry of the ESP encounter, the impact parameter $\mu$ and consequently $\Delta N$ are only sensitive to changes of the initial field values perpendicular to the trajectory.   
Note that the perpendicular component of the inflaton plays  
a similar r\^ole as the modulaton field introduced in \cite{Langlois:2009jp}.

Hence we can write the power-spectrum as a sum of two contributions, the usual slow--roll contribution and a new one caused by the ESP encounter,
\begin{equation}
\mathcal P_{\zeta}\equiv\mathcal P_{\mbox{\tiny{SR}}}+\mathcal P_{\mbox{\tiny{ESP}}}=\left(\frac{H_*}{2\pi}\right)^2\left[\frac{1}{2\varepsilon^*}+
 \Delta N_{,\perp}^2 \right]\,. \label{power-spectrum}
\end{equation}
where $\varepsilon^*=-\dot{H}/H^2$ is the usual slow-roll parameter. 
Here $\Delta N$ is comprised of two contributions that decouple at first order: one due to the slowing-down (subscript ``sl'') of the fields and the other due to the geometric change of the trajectory (subscript ``ge''), 
\begin{eqnarray}
\Delta N=\Delta N_{sl}+\Delta N_{ge} \,.
\end{eqnarray}
Defining 
\begin{eqnarray}
\Xi\equiv \frac{\mathcal{P}_{\mbox{\tiny ESP}}}{\mathcal{P}_\zeta}\,, \label{Xi}
\end{eqnarray}
the scalar spectral index is given by \cite{Wands:2002bn}
\begin{eqnarray}
n_s-1\equiv \frac{d \ln (P_{\zeta})}{ d\ln (k)} =-(6-4\Xi)\varepsilon^*+2(1-\Xi)\eta_{\parallel}^*+2\Xi\eta_{\perp}^*. 
\label{ns-1}
\end{eqnarray}
where $\eta_{\parallel}^*$ and $\eta_{\perp}^*$ correspond to 
 the usual slow-roll parameters parallel and perpendicular to the trajectory evaluated sixty efoldings before the end of inflation. For $\Xi\rightarrow 0$, the slow roll result is recovered, while for larger $\Xi$ the model is similar to the mixed inflaton/curvaton scenario \cite{Langlois:2004nn} or modulated traping \cite{Langlois:2009jp}\footnote{In this reference, $\eta_{\perp}^*\ll \eta_{\parallel}^*$ was assumed. For us $\eta_{\perp}^*=\eta_{\parallel}^*$, due to (\ref{sqpot}).}, see forward to Sec.~\ref{sec:tunedESP}. The tensor--to--scalar ratio can be computed to \cite{Wands:2002bn}
\begin{eqnarray}
r=16 \varepsilon^* (1-\Xi) \,.\label{scalartotensor}
\end{eqnarray}

\subsubsection{The bi- and trispectrum}
One can use  the $\delta N$-formalism to compute the non-Gaussianities of the curvature perturbation associated with the trapping effect. 
Let us first recall the definition of  the local non-linearity parameters $f_{NL}$ (bispectrum) as well as $\tau_{NL}$ and $g_{NL}$ (trispectrum). 
For a multi-field system, the expansion of the number of efoldings in terms of the scalar field fluctuations can be written as 
\beq
\zeta=\delta N= N,_{i} \, \delta \varphi^i + \frac{1}{2}
 N,_{ij} \, \delta \varphi^i \, \delta \varphi^j+\frac16 N,_{ijk} \, \delta \varphi^i \, \delta \varphi^j \,\delta\varphi^k+\dots,
\eeq
where we use the implicit summation convention for the field indices $i, j, k, \dots$ and the notation
 $N_{i}\equiv \partial N/\partial \varphi^i|_*$, $N,_{ij}\equiv \partial^2 N/\partial\varphi^i\partial\varphi^j|_*$, etc. 
 
If all of the scalar field fluctuations are Gaussian, with the power-spectrum (\ref{Pperp}), one finds, using Wick's theorem, that the bispectrum, i.e. the three-point function in Fourier space, is given by
\begin{eqnarray}
\label{bispectrum}
\langle \zeta_{\bk_1} \zeta_{\bk_2} \zeta_{\bk_3} \rangle &\equiv& (2 \pi)^3
\delta^{(3)}\left(\sum_i \bk_i\right) 
B_\zeta (\bk_1,\bk_2,\bk_3),\cr
B_\zeta (\bk_1,\bk_2,\bk_3)&=&\frac65f_{\rm NL}\left[P(k_1)P(k_2)+P(k_2)P(k_3)+P(k_3)P(k_1)\right], \end{eqnarray}
with the non-linearity parameter
\beq
\label{f_NL_gen}
\frac{6}{5}f_{\rm NL} = 
  \frac{
N,_{i} N,_{j} N^{,ij}}{( N,_{k}N^{,k})^2}\,.
\eeq
Similarly, the trispectrum, defined as 
\beq
\langle \zeta_{\bk_1} \zeta_{\bk_2} \zeta_{\bk_3} \zeta_{\bk_4} \rangle_{c} \equiv (2 \pi)^3
\delta^{(3)}\left(\sum_i \bk_i\right) 
T_\zeta (\bk_1, \bk_2, \bk_3, \bk_4)\,,
\eeq
can be written in the form~\cite{Byrnes:2006vq}
\beq
\label{trispectrum}
T_\zeta (\bk_1, \bk_2, \bk_3, \bk_4)=\tau_{\rm NL}\left[P(k_{13})P (k_3)P(k_4)+ 11 \ {\rm perms}\right]
+\frac{54}{25} g_{\rm NL}\left[P(k_2)P(k_3)P(k_4)+3\ {\rm perms}\right],
\eeq
with 
 \beq
  \tau_{\rm NL}= \frac{N_{,ij}N^{,ik}N^{,j}N_{,k}}{(N_{,l}N^{,l})^3}\qquad , \qquad
  g_{\rm NL}=\frac{25}{54}\frac{N_{,ijk}N^{,i}N^{,j} N^{,k}}{(N_{,l}N^{,l})^3}
  \eeq
and where $k_{13}\equiv\|\bf {k}_1+\bf {k}_3\|$.

In the present case, one finds that the non-linearity parameters due to the modulated trapping are given by 
\begin{eqnarray}
f_{NL}&=&\frac{5}{6}\frac{\Delta N_{,\perp}^2\Delta N_{,\perp\perp}}{(\Delta N_{,\perp}^2+N_{,\parallel}^2)^2}=\frac56\Xi^2\, \frac{\Delta N_{,\perp\perp}}{(\Delta N_{,\perp})^2}\,,\\
\tau_{NL}&=&\frac{\Delta N_{,\perp}^2\Delta N_{,\perp\perp}^2}{(\Delta N_{,\perp}^2+N_{,\parallel}^2)^3}= \frac{(\Delta N_{,\perp\perp})^2}{(\Delta N_{,\perp})^4}\ \Xi^3=\frac{36}{25}\, \Xi^{-1}\, f_{\rm NL}^2\,,\\
g_{NL}&=&\frac{25}{54}\frac{\Delta N_{,\perp}^3\Delta N_{,\perp\perp\perp}}{(\Delta N_{,\perp}^2+N_{,\parallel}^2)^3}=\frac{25}{54}\frac{\Delta N_{,\perp\perp\perp}}{(\Delta N_{,\perp})^3}\ \Xi^3\,.
\end{eqnarray} 
The inequality that $\tau_{NL}\geq(6f_{NL}/5)^2$, which here follows from the fact that $\Xi\leq1$, is true much more generally \cite{Suyama:2007bg}. However it might be violated in special models in which the ``loop" corrections are very large \cite{Sugiyama:2011jt}.

\subsection{Slowing down \label{sec:slowingdown}}
Our next goal is to estimate $\Delta N_{sl}$ analytically, closely following  \cite{Langlois:2009jp}. Assume that after the ESP encounter, the change in $\varphi_\perp$ remains small compared to $\varphi_\parallel$ and $\mu$ and consider that $|\varphi_\parallel-\varphi_p| \gg \mu$. In this case, we can approximate the equation of motion (\ref{eom}) for $\varphi_\parallel$ by (see (\ref{eq1}) in the appendix) 
   \begin{equation}
\ddot\varphi_\parallel+3H_p\dot\varphi_\parallel+\frac{\partial V}{\partial\varphi_\parallel}=gn_p\left(\frac{a_p}{a}\right)^{3}\Theta(t-t_p)\,.\label{eom0}
\end{equation}
We would like to solve this equation over a brief time interval of a few Hubble times only; in this case we can treat $H\approx H_p=\mbox{const}$ and integrate (\ref{eom0}) to
\begin{eqnarray}
\dot{\varphi}_{\parallel}\approx -v_p +gn_pe^{-3H_p(t-t_p)}(t-t_p)\,,
\end{eqnarray}
were we imposed $|\dot\varphi_\parallel|=v_p$ at $t_p$. By denoting
\begin{equation}
\Delta\varphi_{\parallel}(t)\equiv\varphi_\parallel(t,g\neq 0)-\varphi_\parallel(t,g=0)\,,
\end{equation}
we get
\begin{equation}
\Delta\dot\varphi_\parallel=gn_pe^{-3H_p(t-t_p)}(t-t_p)\,, \label{velocitydecrease}
\end{equation}
which gives the amount by which the field is slowed down after the ESP encounter. 
Thus, the trapping event delays the field by
\begin{equation}
\Delta\varphi_\parallel=\int_{t_p}^{\infty}\Delta\dot\varphi_\parallel dt=\frac{gn_p}{9H_p^2}\,,\label{Deltavarphiparallel}
\end{equation}
and the net change in the number of efoldings becomes
\begin{eqnarray}
\Delta N_{sl}&\approx&-\frac{H_p\Delta\varphi_\parallel}{\dot\varphi_*}\\
&=&\frac{g^{5/2}v_p^{1/2}}{9H_p(2\pi)^3}e^{-\pi g\frac{\mu^2}{v_p}} \label{DeltaNsl}
\,.
\end{eqnarray}
Note the presence of the exponential, which is absent in the one dimensional case.
 
A small change in the position of the inflaton at time $t_*$ induces a variation of the angle $\beta$ between the inflationary trajectory and the $\varphi_1$-axis (see Fig.~\ref{pic:trajectory1}), which entails a change of the impact parameter
\begin{eqnarray}
\delta\mu \approx \varphi_p \delta \beta\approx  \frac{\varphi_p}{\varphi_*}\varphi_{\perp}^*\,. \label{muandphistar}
\end{eqnarray}
In this regime, the power-spectrum (\ref{power-spectrum}) picks up a piece that is boosted by
\begin{equation}
\Delta N_{sl,\perp}^2\approx\left(\frac{\varphi_p}{\varphi_*}\right)^2\frac{1}{H_p^2}\frac{g^7}{v_p}\frac{\mu^2}{(6\pi)^4}e^{-2\frac{\pi g\mu^2}{v_p}}\label{deltaNslperp}
\equiv A^2C^2xe^{-2x}\,,
\end{equation}
where we have defined
\begin{eqnarray}
A\equiv \frac{g^{5/2}\sqrt{v_p}}{H_p9(2\pi)^3}\,\,\,,\,\,
C\equiv 2\frac{\varphi_p}{\varphi_*}\sqrt{\frac{\pi g}{v_p}}\,\,\,,\,\,
x\equiv \frac{\pi g\mu^2}{v_p}\,. \label{ACx}
\end{eqnarray}

Moreover,  the additional contributions to the non-linearity parameters due to the slowing down effect become
\begin{eqnarray}
f_{NL}^{sl}&\approx&\frac{5}{6}\frac{A^3C^4x\left(x-\frac{1}{2}\right)e^{-3x}}{\left(A^2C^2xe^{-2x}+\frac{\phi_*^2}{4}\right)^2}\,, \label{fNLsl}\\
\tau_{NL}^{sl}&\approx&\frac{A^4C^6x\left(x-\frac{1}{2}\right)^2e^{-4x}}{\left(A^2C^2xe^{-2x}+\frac{\phi_*^2}{4}\right)^3}\,,\label{tauNLsl}\\
g_{NL}^{sl}&\approx&\frac{25}{54}\frac{A^4C^6x^2\left(x-\frac{3}{2}\right)e^{-4x}}{\left(A^2C^2xe^{-2x}+\frac{\phi_*^2}{4}\right)^3}\,.\label{gNLsl}
\end{eqnarray}
Here, we also assumed that effects due to the geometric change of the trajectory are much smaller than the ones caused by slowing down, see Sec.\ref{sec:geom}.

Under which conditions can we trust these analytic expressions and when are they observable? The 
additional contribution to the
power-spectrum  
is maximal at
\begin{equation}
\mu_{c}=\sqrt{\frac{v_p}{2\pi g}}\,.\label{muc}
\end{equation}
with a corresponding amplitude of
\begin{equation}
\mathcal P_{sl}(\mu_{c})\simeq \frac{g^6}{e(2\pi)^73^4}\approx 1.2\times 10^{-8}g^6\,,
\end{equation}
where we used the slow--roll approximation $\varphi_pH_*/(\varphi_*H_p)\simeq 1$ as well as $\beta\ll 1$ 
(see Fig.~\ref{pic:trajectory1}).
Given a large enough coupling $g$, it is possible that this contribution dominates, saturating the COBE bound. To find the coupling $g_{d}$ 
beyond
 which the trapping effect can dominate we solve $\mathcal P_{sl}/\mathcal P_{\mbox{\tiny{SR}}}=1$ 
to
\begin{equation}
g_{d}=(m\varphi_{*}^2)^{1/3}\left((2\pi)^5e\right)^{1/6}\left(\frac{3}{2}\right)^{1/2}\approx 3.9 (m\varphi_{*}^2)^{1/3}\,. \label{gd}
\end{equation}
If $g>g_d$, there exists a range around $\mu_c$ for which the amplitude is larger than the usual slow--roll contribution. Of course it is then this contribution which needs to be tuned, i.e. by carefully choosing $\varphi_\perp^*$. We investigate this model, including a discussion of non-Gaussianities, in Sec.~\ref{sec:tunedESP}.

If the coupling becomes too large, a qualitative change in the trajectory occurs: instead of being merely slowed down, the velocity reverses and the trajectory gets temporarily trapped. This takes place when the force due to backreaction is bigger than the one stemming from the classical potential, 
\begin{equation}
\left|\frac{\partial V}{\partial \varphi_1}\right|_p < \left|\left(\frac{\rho_{\chi}}{\partial\varphi_1}\right)\right|_p\simeq \frac{g^{5/2}v_p^{3/2}}{(2\pi)^3}e^{-\frac{\pi g\mu^2}{v_p}}\,,
\end{equation}
where the subscript $p$ refers to evaluating a quantity at the time of particle production $t_p$. Solving for $g$ in the $\mu=0$ case yields
\begin{equation}
g_{\mbox{\tiny{trap}}}\equiv (m\varphi_{p}^2)^{1/5}\left((2\pi)^3\left(\frac{3}{2}\right)^{3/4}\right)^{2/5}\approx 3.2 (m\varphi_{p}^2)^{1/5}\,, \label{gtrap}
\end{equation}
which is bigger than the coupling at which the power-spectrum can be dominated by $\mathcal{P}_{sl}$
\footnote{$\frac{g_d}{g_{trap}}\sim \frac{\varphi_*^{2/3}m^{2/15}}{\varphi_p^{2/5}}\sim \varphi_p^{-2/5}<1$ for $\varphi_*=16$ and $m=10^{-6}$; since inflation ends for $1.4\simeq \varphi < \varphi_p <\varphi_*$ and the inequality becomes stronger for smaller $m$ we get $g_{\mbox{\tiny trap}}>g_d$.}.

However, even for large $g>g_{\mbox{\tiny{trap}}}$, the slowing down effect is well approximated by the above analytic expression if the impact parameter is bigger than 
\begin{eqnarray}
\mu>\mu_{an}\equiv \sqrt{\frac{m}{g\pi}\sqrt{\frac{2}{3}}\ln\left(\frac{g^{5/2}}{m\varphi_p}\left(\frac{2}{3}\right)^{3/4}\frac{1}{(2\pi)^3}\right)}\,.\label{muan}
\end{eqnarray}
This parameter is usually bigger than the one for which the power-spectrum has its peak, but low enough to allow for the use of the analytic approximation in Sec.~\ref{sec:tunedESP}.

\subsection{Geometric change of the trajectory \label{sec:geom}}
In this section we would like to estimate the path change compared to the trajectory in the absence of an ESP. The trajectory bends towards the ESP after particle production took place.  Our goal is to estimate the maximal deviation in the perpendicular direction $\Delta\varphi_{\perp}$ under the assumption that $\Delta\varphi_{\perp}\ll\mu$, which in turn leads to a slight enhancement of the number of efoldings $\Delta N_{\perp}$, in addition to the slowing down effect discussed in the previous section. Since 
$\mu>\Delta \varphi_\perp$ (see Fig.\ref{pic:trajectory2}), the effect is small compared to the slowing down effect. It becomes much stronger for large coupling $g>g_{\mbox{\tiny{trap}}}$ and small impact parameters $0<\mu<\mu_{an}$, for which it is challenging to find analytic results. 
We compute such cases numerically in Sec.~\ref{sec:numerics}, enabling a comparison of our analytic approximations to the full solution.

There are two different regimes for $\varphi_{\perp}\ll \mu$: in the first one, back-reaction is strong and the trajectory bends, which requires $|\varphi_\parallel-\varphi_p|\sim v_p (t-t_p) \ll \mu$, and we can approximate the equation of motion (\ref{eom}) by (for ease of notation, we put the ESP above the trajectory in this section $\varphi_{\perp\mbox{\tiny ESP}}=+\mu$ with $\mu>0$, opposite to Fig.~\ref{pic:trajectory1} and Fig.~\ref{pic:trajectory2} but in line with App.~\ref{app:2}, i.e.~eqn.~(\ref{eq2}))
\begin{eqnarray}
\label{eom_perp}
\ddot{\varphi}_{\perp}+3H_p\dot{\varphi}_{\perp}+m^2\varphi_{\perp}\approx gn_pe^{-3H_p(t-t_p)}\,,
\end{eqnarray}
where we treat $H_p\approx \mbox{const}$, since we are only interested in following $\varphi_{\perp}$ until $v_p(t-t_p)=\mu$ (a fraction of an efolding). Ignoring the classical potential ($m=0$), the above can be integrated to
\begin{eqnarray}
\varphi_\perp(t)=\frac{g n_p}{9H_p^2}\left(1-e^{-3H(t-t_p)}(1+3H_p(t-t_p))\right)\,.
\end{eqnarray}
Hence, the excursion at $\tilde{t}-t_p=\mu/v_p$ is of order  
\begin{eqnarray}
 \varphi_\perp(\tilde{t}) \sim \frac{g n_p}{9H_p^2}\frac{9}{8}\mu^2\varphi_p^2 =\Delta\varphi_{\parallel}\frac{9}{8}\mu^2\varphi_p^2\,, \label{Deltavarphiperp}
\end{eqnarray}
with $n_p$ from (\ref{n_p}), we expanded the exponential for small $v_p\mu/H\simeq 2\mu/\varphi_p\ll 1$ and we used
$\Delta \varphi_{\parallel}$ from (\ref{Deltavarphiparallel}). We see that $\Delta \varphi_\perp \ll \mu$ for small $\mu$.  The corresponding velocity in the $\perp$-direction is approximately 
\begin{eqnarray}
v_\perp\equiv \dot{\varphi}_\perp(\tilde{t})\approx \frac{\varphi_p \mu g n_p}{2H_p}\,.
\end{eqnarray}

Once $v_p(t-t_p)>\mu$, the source term on the right hand side of (\ref{eom}) becomes unimportant quickly and the second phase commences: the field rolls freely until the classical potential and Hubble friction bring the trajectory back onto a slow--roll attractor. This takes place roughly within one Hubble time, see Fig.~\ref{pic:trajectories}, panel a. The total excursion in the perpendicular direction can be estimated by solving
\begin{eqnarray}
\ddot{\varphi}_{\perp}+3H_p\dot{\varphi}_{\perp}\approx 0
\end{eqnarray}
with $\varphi_\perp^{ini}\approx 0$ (we checked that using $\varphi_\perp$ from (\ref{Deltavarphiperp}) does not affect results significantly) and $\dot{\varphi}_{\perp}^{ini}=v_\perp$ from above to
\begin{eqnarray}
\varphi_\perp(t)\approx \frac{v_\perp}{3H_p}\left(1-e^{-3H_p(t-t_p)}\right)\,,
\end{eqnarray}
and taking the large-$t$ limit (which maximises $\varphi_\perp(t)$ in the limit that $m_\perp=0$), we find
\begin{eqnarray} 
\Delta\varphi_\perp\approx \Delta \varphi_\parallel\mu\varphi_p\frac{3}{2}\,.\label{Deltavarphiperp2}
\end{eqnarray}
Thus, the trajectory is prolonged by roughly\footnote{An alternative viewpoint might be instructive: during the first phase, the trajectory is bent by the angle $\tan(\alpha)=\varphi_\perp(\tilde{t})/\mu$, with respect to the unperturbed trajectory; since this angle is small we have $\alpha\sim \Delta\varphi_{\parallel}\mu\varphi_p^29/8$. Since the overall speed decreases in a grazing ESP encounter and the decrease is not strong in the cases of interest, see Fig.~\ref{pic:trajectories} panel b, we take the slow--roll speed $v_p$ as an approximation for the speed along the intermediate trajectory before a slow--roll attractor is reached again (this provides an upper bound on effects). Then the difference in path-length during the excursion is of order 
$\Delta s\sim v_p/H_p-l,$ where $l=v_p\cos(\alpha)/H_p\approx (1-\alpha)v_p/H_p$. Plugging in $\alpha$ gives
$\frac{\Delta s}{\Delta \varphi_{\parallel}}\sim \frac{9}{4}\mu\varphi_p$, consistent (up to factors of order one) with the above estimate.}
\begin{eqnarray}
\Delta s\sim \Delta\varphi_\perp\approx \Delta \varphi_\parallel\mu\varphi_p\frac{3}{2}\,,
\end{eqnarray} 
and the corresponding change in the number of efoldings can be estimated, up to factors of order unity, to
\begin{eqnarray}
\Delta N_{ge}&\sim& \frac{H_p\Delta s}{v_p} \,.
\end{eqnarray}
For small impact parameters, this change in the number of efoldings is much smaller than the corresponding one from slowing down,
\begin{eqnarray}
\frac{\Delta N_{ge}}{\Delta N_{sl}}\sim \frac{3\varphi_p  \mu}{2}\,.
\end{eqnarray}
Hence, as long as $\mu\ll 2/(\varphi_p 3)$, slowing down effects are dominant. Furthermore, since the ratio of the derivatives is 
\begin{eqnarray}
\label{ratio_sl_ge}
\frac{\Delta N_{ge,\perp}}{\Delta N_{sl,\perp}}\sim -\frac{3 \mu\varphi_p }{2}\left(\frac{v}{2\pi g \mu^2}-1\right)\,,
\end{eqnarray}
 the slope  is dominated by the slowing down effects for $\mu_{an}<\mu \ll 2/(3\varphi_p )$  where $\mu_{an}=\mbox{few}\times\sqrt{v_p/(\pi g)}$ (the same holds true for higher--order derivatives up to factors of order unity). For small $\mu$ and small $g$ the geometric effects can be bigger than slowing down effects at the level of correlation functions; however, in this regime the usual slow--roll contributions tend to dominate, since corrections scale as $\mu$. This is the reason why, for $\mu> \mu_{an}$, we are justified to ignore $\Delta N_{ge}$ in Sec.~\ref{sec:slowingdown} when computing the power-spectrum and the non-linearity parameters. Then, the additional contribution  to the power-spectrum  and higher-order correlation functions caused by the geometric change in the trajectory are of order
\begin{eqnarray}
\mathcal{P}_{ge}&\sim & \left(\frac{H_*}{2\pi}\right)^2A^2B^2C^2e^{-2x}\left(x-\frac{1}{2}\right)^2\,,\\
f_{NL}^{ge}&\sim & \frac{5}{6}\frac{A^3B^3C^4x^{1/2}e^{-3x}\left(x-\frac{1}{2}\right)^2\left(x-\frac{3}{2}\right)}{\left(A^2C^2xe^{-2x}+\frac{\phi_*^2}{4}\right)^2}\,,\label{fNLge}\\
\tau_{NL}^{ge}&\sim & \frac{A^4B^4C^6xe^{-4x}\left(x-\frac{3}{2}\right)^2\left(x-\frac{1}{2}\right)^2}{\left(A^2C^2xe^{-2x}+\frac{\phi_*^2}{4}\right)^3}\,,\\
g_{NL}^{ge}&\sim & \frac{25}{54}\frac{A^4B^4C^6e^{-4x}\left(x^2-3x+\frac{3}{4}\right)\left(x-\frac{1}{2}\right)^3}{\left(A^2C^2xe^{-2x}+\frac{\phi_*^2}{4}\right)^3}\,,
\end{eqnarray}
where we defined
\begin{eqnarray}
B\equiv \frac{3\varphi_p}{2}\sqrt{\frac{v_p}{g\pi}}\,. \label{B}
\end{eqnarray}
 If $g<g_{\mbox{\tiny trap}}$, so that the analytic approximations are valid for all $\mu$, and $\mu \ll \mu_{an}$, so that $|\Delta N_{,\perp}^{sl}|\ll |\Delta N_{,\perp}^{ge}|$, one only needs to replace the denominators above by appropriate powers of $(C^2A^2B^2e^{-2x}(x-1/2)^2+\varphi_*^2/4)$. Note that $\Delta N_{,\perp}^{ge}$ changes sign for $\mu<\sqrt{v/(2\pi g)}$, indicating that increasing the impact parameter from zero up first leads to an increase in the number of efoldings, in accord with the trajectory being bent, before the weakening of the slowing down effect decreases $N_{total}$ and, for $\mu>\sqrt{v/(2\pi g)}$, trajectories show less of an excursion again, see Fig.~\ref{pic:trajectories} (a).

\subsection{Discussion and numerical studies \label{sec:numerics}}
\begin{figure}[tb]
\begin{center}
\includegraphics[scale=0.6,angle=0]{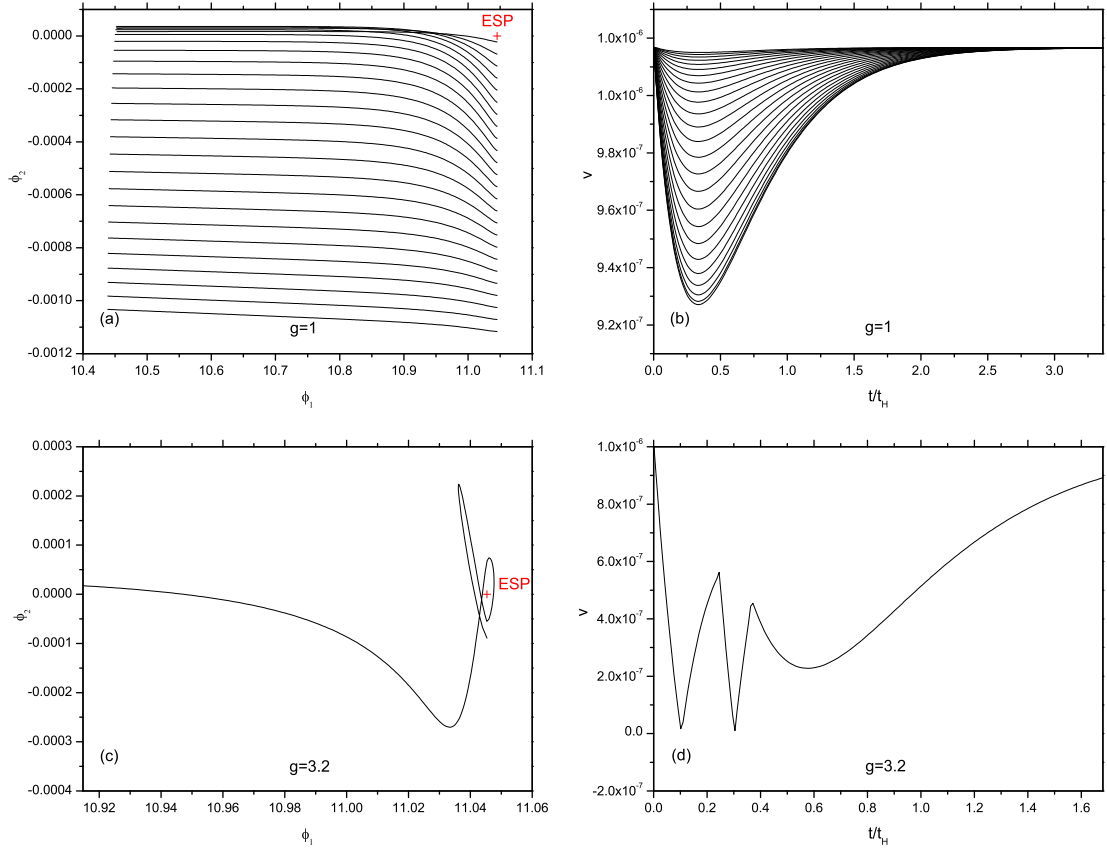}
   \caption{\label{pic:trajectories} (a) and (b) show the trajectory and corresponding velocity $v=\sqrt{\dot{\varphi}_1^2+\dot{\varphi}_2^2}$ (over $t/t_H\approx t m\varphi_{p}/\sqrt{6}$) directly after an ESP encounter ($\chi$ particles are produced instantaneously at $t=0$ only) with $g=1<g_{\mbox{\tiny trap}}\approx 1.8$ for a range of impact parameters $\mu$ (the same as in Fig.~\ref{pic:efoldsovermu} (b)). The field slows down while the trajectory gets slightly disturbed, panel (a). The slowing down effect, which dominates correlation functions, is monotonically decreasing with increasing $\mu$, and $v$ approaches the slow--roll speed again after a few Hubble times, panel (b).  (c) and (d) show a temporarily trapped trajectory with corresponding velocity, with an increased coupling $g=3.2>g_{\mbox{\tiny trap}}$.   The fields oscillate temporarily around a non-parabolic minimum of the effective potential, and the resulting trajectories are highly sensitive to initial conditions. This leads to a non-analytic dependence of the number of efoldings on initial conditions for small impact parameters, Fig.~\ref{pic:efoldsovermu} (d). In all plots $m=0.2\times m_{\mbox{\tiny COBE}}$, defined in (\ref{mCOBE})    
   (larger $m$ increase $g_{\mbox{\tiny trap}}$). }
   \end{center}
\end{figure}

We encountered several different regimes with qualitatively different behaviour, see table \ref{tab:regimes} for  a summary. All regimes can be treated  numerically without invoking the slow--roll approximation (which can be violated during the ESP encounter) or the small $\mu$ approximation, as long as we assume a single, instantaneous particle production event. The latter can be a good approximation even for temporarily trapped trajectories, since the velocity decreases after the first encounter with the ESP, $v<v_p$, and particle production is less efficient subsequently, (\ref{n_p}) (see Fig.~\ref{pic:trajectories} (c) and (d) for a concrete case).

\begin{table}[tb]

\begin{center}
\begin{tabular}{llll}
\hline\hline
Coupling Constant &Impact Parameter & Description\\
\hline
$g< g_{\mbox{\tiny COBE}}$ in (\ref{gCOBE}) & $\mu=\mbox{any}$ & $\mathcal{P}_{\mbox{\tiny COBE}}>\mathcal{P}_{\mbox{\tiny ESP}}$   \\
\hline
$g_d<g$ in (\ref{gd})& $\mu$ near $\mu_c$ in (\ref{muc})& $\mathcal{P}_{\mbox{\tiny infl}}<\mathcal{P}_{\mbox{\tiny ESP}}$    \\
\hline
$g<g_{\mbox{\tiny max}}$ in (\ref{gmax})& $\mu=\bar{\mu}$ in (\ref{barmu})& The non-adiabaticity parameter in (\ref{adiabaticity}) is bigger than one\\
& & and particle production is fast (this paper).\\
$g\gg g_{\mbox{\tiny max}}$ & $\mu=\bar{\mu}$ in (\ref{barmu})& The approximation in App.~C of \cite{Kofman:2004yc} could be used,\\
&& but NG are expected to be too large for $m\sim m_{\mbox{\tiny COBE}}/\mbox{few}$.\\
\hline
 $g_d<g<g_{\mbox{\tiny trap}}$ in (\ref{gtrap}) & $\mu \ll 2/(\varphi_p 3)$ & Slowing down effects dominate over geometric effects. \\
                                 & $\mu=\bar{\mu}$ & If $m<m_{\mbox{\tiny COBE}}$ in (\ref{mCOBE}) then $\mathcal{P}_{\mbox{\tiny COBE}}$ is saturated by $\mathcal{P}_{sl}$ at $\bar{\mu}$. \\
                         & $\mu \gtrsim 2/(\varphi_p 3)$ & Geometric effects dominate over slowing down effects, \\
                          &  &  but they are usually negligible compared to slow-roll effects.  \\
                          \hline
 $g_{\mbox{\tiny trap}}<g$  & $\mu<\mu_{an}$ in (\ref{muan}) & Temporary trapping can occur,  \\
 & & $N(\mu)$ can become non-analytic, Fig.~\ref{pic:efoldsovermu} (d).\\
                        & $\mu_{an}<\mu \ll 2/(\varphi_p 3)$ & No trapping, slowing down effects dominate. \\
                                 & $\mu=\bar{\mu}$& If $m<m_{\mbox{\tiny COBE}}$ then $\mathcal{P}_{\mbox{\tiny COBE}}$ is saturated by $\mathcal{P}_{sl}$ at $\bar{\mu}$. \\
                         & $\mu \gtrsim 2/(\varphi_p 3)$ & Geometric effects dominate over slowing down effects, \\
                            &  &  but they are usually negligible compared to slow-roll effects.  \\
                            \hline
$0.8<g<2.5$   &$\mu=\bar{\mu}$& $\mathcal{P}_{\mbox{\tiny COBE}}$ is matched by $\mathcal{P}_{sl}$ in Sec.~\ref{sec:tunedESP}  and non-Gaussianities \\
& & are observable (we use $0.02<m/m_{\mbox{\tiny COBE}}<0.2$). \\
\hline\hline
\end{tabular}
\label{tb:}
\caption{Different regimes depending on the coupling constant $g$ and the impact parameter $\mu$.\label{tab:regimes}}
\end{center}
\end{table}

\begin{figure}[tb]
\begin{center}
\includegraphics[scale=0.6,angle=0]{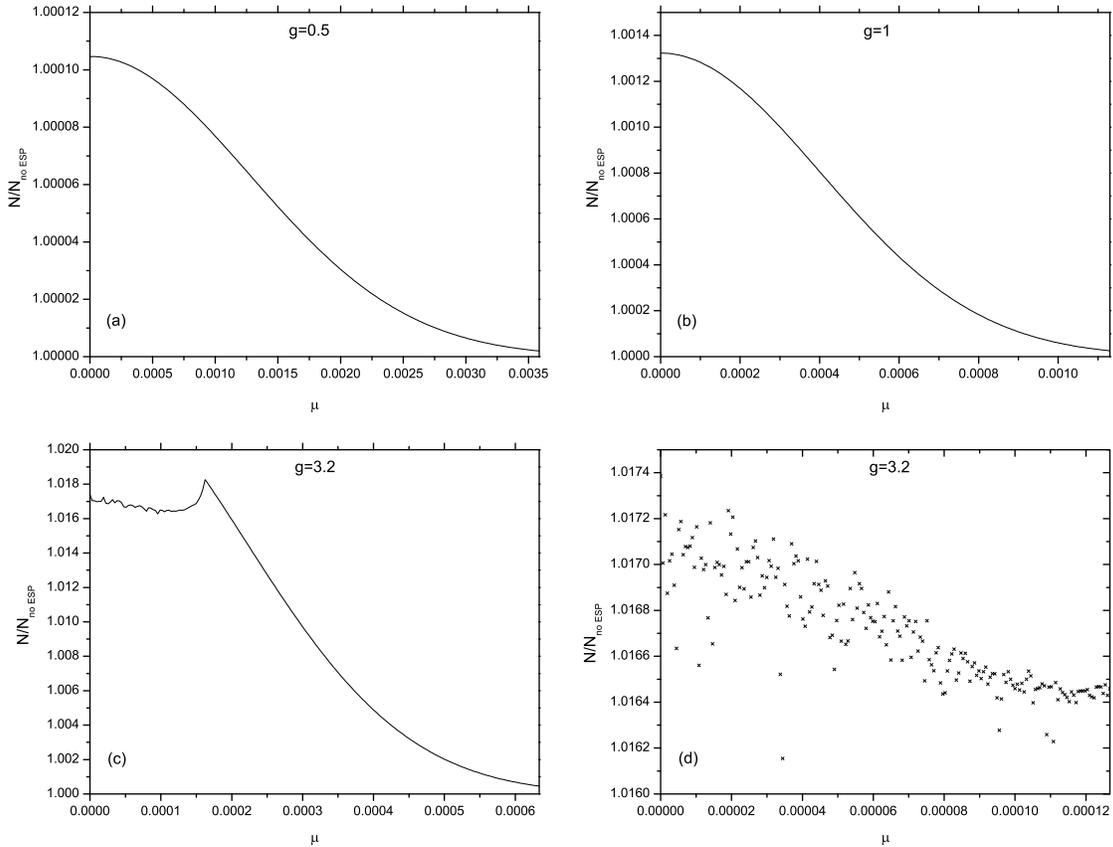}
   \caption{\label{pic:efoldsovermu} The total number of efoldings $N$ in the presence of an ESP divided by the number of efoldings in the absence of on ESP, $N_{\mbox{\tiny no ESP}}$, is plotted over the impact parameter $\mu$, for (a) $g=0.5$, $m=m_{\mbox{\tiny COBE}}$ (slow--roll contributions dominate the power-spectrum), (b) $g=1$, $m=0.2\times m_{\mbox{\tiny COBE}}$ (the slowing down effect can dominate the power-spectrum), (c) and (d) $g=3.2$, $m=0.2\times m_{\mbox{\tiny COBE}}$ (for small $\mu$ the trajectory becomes temporarily trapped, the system becomes chaotic and $N(\mu)$ non-analytic, see panel (d); for $\mu>\mu_{an}$, trapping does not occur and the slowing down effect dominates again).   }
   \end{center}
\end{figure}

\begin{figure}[tb]
\begin{center}
\includegraphics[scale=0.6,angle=0]{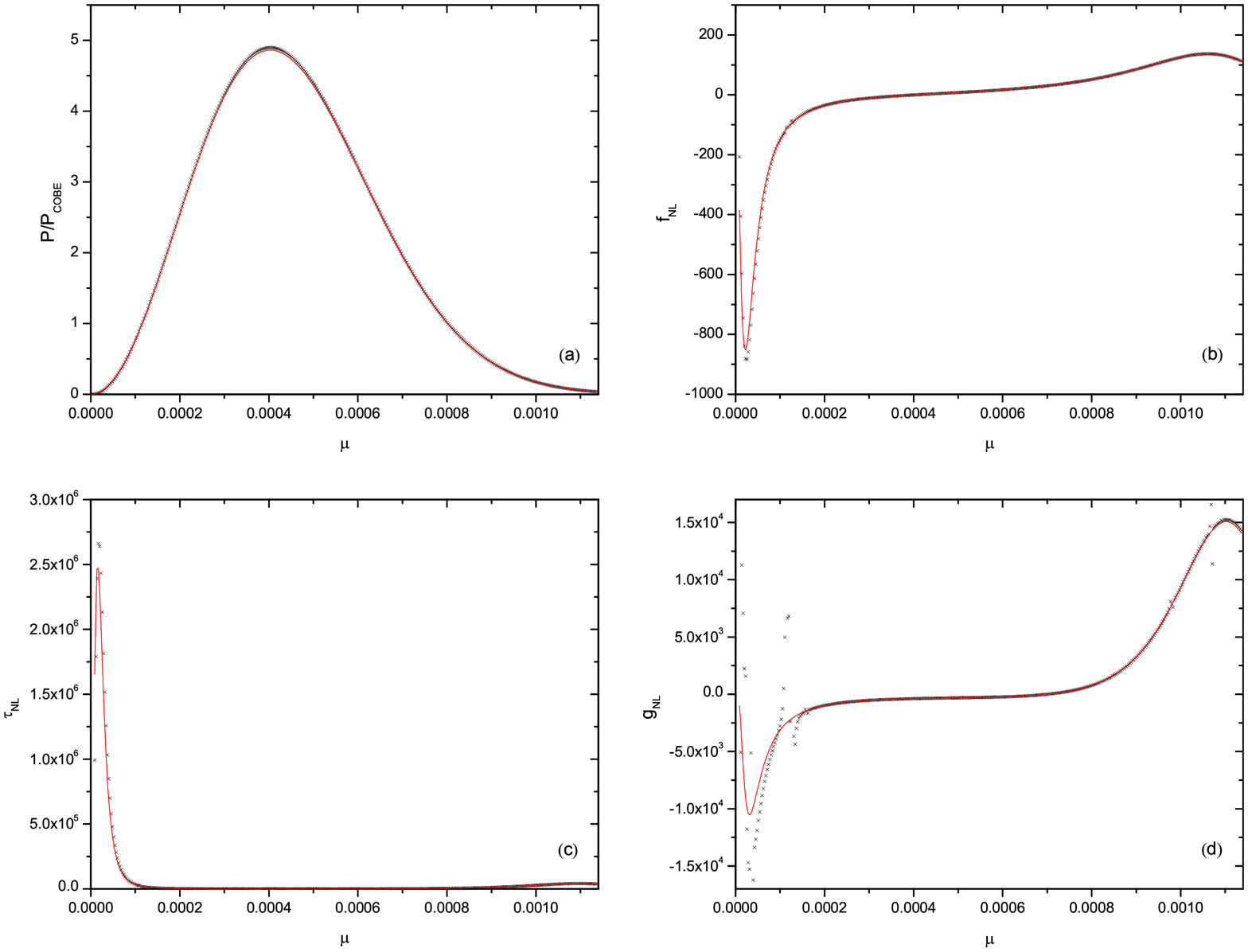}
   \caption{\label{pic:PandNLnumvsan}  The additional contribution to the power-spectrum normalized by $\mathcal{P}_{\mbox{\tiny COBE}}$ as well as the non-linearity parameters $f_{NL}$, $\tau_{NL}$ and $g_{NL}$ are computed numerically (crosses) and compared to the analytic approximations (red line) in Sec.~\ref{sec:slowingdown} which lead to (\ref{deltaNslperp}) and (\ref{fNLsl})-(\ref{gNLsl}) using $g=1$ and $m=0.2\times m_{\mbox{\tiny COBE}}$; slow--roll contributions are negligible and as a consequence, the impact parameter $\mu$ needs to be tuned in order to satisfy the COBE bound (the normalized power-spectrum in (a) needs to equal one, singling out two values of $\mu$), see forward to Sec.~\ref{sec:tunedESP}. The discrepancy between numeric and analytic results for large and small $\mu$ are due to additional effects caused by the geometric change of the trajectory, see Sec.~\ref{sec:geom}, which are negligible in Sec.~\ref{sec:tunedESP} where  $\mathcal{P}/\mathcal{P}_{\mbox{\tiny COBE}}=1$ at the larger of the two possible impact parameters.  Rounding errors amplify quickly once $\mathcal{P}/\mathcal{P}_{\mbox{\tiny COBE}}\ll 1$ (negligible in Sec.~\ref{sec:tunedESP}). } 
   \end{center}
\end{figure}

First, we would like to investigate some trajectories in field space and how they compare to our expectation. In Fig.~\ref{pic:trajectories} (a) we can see the that the bending of a trajectory grows as $\mu$ is increased from zero, before they remain essentially straight for larger impact parameters. This is in line with our predictions in Sec.~\ref{sec:geom}, where we observed a sign change in the second derivative of $\Delta N_{ge}$ in (\ref{fNLge}). The corresponding velocity change is plotted in panel (b). At the particle production event, only a small fraction of the energy is transferred to $\chi$ particles, less than 
$1\%$ in Fig.~\ref{pic:trajectories} (b), but the velocity decreases subsequently due to backreaction of $\chi$ particles. As the universe expands, $\rho_\chi$, and thus the force due to backreaction, redshifts as $a^{-3}$, causing the field to speed up again and approach the slow--roll speed $v_{p}$ in (\ref{vp}) after a few Hubble times. The maximal decrease of the velocity is monotonically decreasing as $\mu$ increases, since the exponential suppression in (\ref{DeltaNsl}) becomes important and as a consequence, $f_{NL}^{sl}$ in (\ref{fNLsl}) is always positive. 

The coupling in panels (a) and (b) of Fig.~\ref{pic:trajectories} satisfies $g<g_{\mbox{\tiny trap}}$, preventing a trapping of trajectories. We chose $g=1$ as a representative coupling, since it is the most natural value for a dimensionless parameter; it is large enough, $g>g_{\mbox{\tiny COBE}}$ (see forward (\ref{gCOBE})) 
and $g_{d}$, for the power-spectrum to be dominated by the slowing down effect, Fig.~\ref{pic:efoldsovermu} (b) and Fig.\ref{pic:PandNLnumvsan} (a). If the coupling is increased above $g_{\mbox{\tiny trap}}\approx 1.8$, trajectories can get temporarily trapped for small impact parameters $\mu<\mu_{an}$. An example of such a trajectory is shown in Fig.~\ref{pic:trajectories} (c) with the corresponding velocity in (d).

If trajectories get trapped \footnote{The way particle production is described as an instantaneous event here and elsewhere \cite{Kofman:2004yc} is not a good approximation to describe trajectories during the actual trapping. Thus, plots like Fig.~\ref{pic:trajectories} (c) and (d) as well as  Fig.~\ref{pic:efoldsovermu} (c) and (d) should be taken with caution. However, the statement that trajectories get trapped and that they are sensitive to initial conditions is robust.}, they oscillate around the minimum of the effective potential close to the ESP.
These trajectories are highly sensitive to initial conditions, as expected for a two dimensional non-harmonic oscillator (the system is chaotic; see \cite{Greene:2007sa} for related work). For example, a trajectory can change qualitatively from having three revolutions to two revolutions at a critical impact parameter. This leads to discrete jumps in the number of efoldings, rendering $N(\mu)$ non-differentiable. For such a chaotic system the $\delta N$ formalism can not be applied. This is evident in Fig.~\ref{pic:efoldsovermu} (c) and (d): for large $\mu$, trajectories are not trapped and the shape of $N(\mu)$ is similar to the ones in panel (a) and (b) where $g<g_{\mbox{\tiny trap}}$, but for small $\mu$ the curve in panel (c) becomes irregular. A zoomed in version of this small $\mu$ range in panel (d) reveals that $N(\mu)$ is not smooth at all. An investigation of chaos in such a system is interesting\footnote{For example, do intermediate regions of smooth $N(\mu)$ exist? What is the fractal dimension of the non-smooth set? Etc. Note that in order to answer these questions, one needs to track subsequent particle production  and the amplification of numerical errors. All numerics in this paper were done with MAPLE9's standard integration routines, $14$ digits accuracy (we varied this accuracy to check that no conclusions in this article are sensitive to it) and a single particle production event.}, but goes beyond the scope of this article. In the following we focus on cases without temporary trapping ($\mu>\mu_{an}$ if $g>g_{\mbox{\tiny trap}}$). A remnant of this sensitivity can be observed in regions where geometric effects become important, see forward to Fig.~\ref{pic:PandNLnumvsan} (d), where a discontinuity in the second derivative of $N$  causes a divergence in $g_{NL}$ for small $\mu$.  

If slowing down effects dominate over geometric effects\footnote{In the case of trapped trajectories, such a separation does not make sense anyhow, since neither of the two analytic descriptions in \ref{sec:slowingdown} or \ref{sec:geom} are applicable.}, $\mu_{an}<\mu \ll 2/(\varphi_p 3)$, $N(\mu)$ is smooth. One can then tune the impact parameter to match the COBE normalization and compute corresponding non-Gaussianities, see Sec.~\ref{sec:tunedESP} and Fig.~\ref{pic:NLftESP}. In order to make analytic predictions, we focus entirely on slowing down effects as described in Sec.~\ref{sec:tunedESP}. This is an excellent approximation, which can be seen in Fig.~\ref{pic:PandNLnumvsan} (a)-(d), where the full numerical solution (crosses) for the power-spectrum and the non-linearity parameters $f_{NL}$, $\tau_{NL}$ and $g_{NL}$ are compared with the analytic approximations (\ref{deltaNslperp}) and (\ref{fNLsl})-(\ref{gNLsl}) for $g=1$ and $m=0.2\times m_{\mbox{\tiny COBE}}$. We chose  $m<m_{\mbox{\tiny COBE}}$ and the most natural coupling $g=1$, since we focus on such cases in Sec.~\ref{sec:tunedESP}. The differences for small and large $\mu$ are caused by geometric effects becoming important. However, it is clear that around the maximum of $\mathcal{P}/\mathcal{P}_{\mbox{\tiny COBE}}$ the analytic approximations are excellent.

\section{Curvature fluctuations from a grazing ESP encounter
\label{sec:tunedESP}}

\subsection{Confronting the model with observations}
If the coupling of the extra species particles is strong, $g>g_d$ in (\ref{gd}), the power-spectrum can be dominated by fluctuations in $\varphi_\perp$ due to the slowing down effect (Sec.~\ref{sec:slowingdown}) provided that the impact parameter is close to $\mu_c$ in (\ref{muc}), but less than $\mu\ll 2/(\varphi_p 3)$ at which point the geometric effect of Sec.~\ref{sec:geom} becomes important, 
according to (\ref{ratio_sl_ge}). 
To satisfy the COBE normalization via the slowing down effect alone, we  need 
\begin{eqnarray}
g>g_{\mbox{\tiny COBE}}\equiv (e(2\pi)^73^4\mathcal{P}_{\mbox{\tiny COBE}})^{1/6}\approx 0.77>g_d\,, \label{gCOBE}
\end{eqnarray}
where $g_{\mbox{\tiny COBE}}$ is defined such that $\mathcal{P}_{sl}(\mu_c)=\mathcal{P}_{\mbox{\tiny COBE}}$.
For $g\gtrsim g_d\sim \mathcal{O}(1)$ and $\varphi_p\sim 10$, the inequality  $\mu\ll 2/(\varphi_p 3)$ implies $\mu\ll 0.1$, which is easy to satisfy. Since $g$ can be larger than $g_{\mbox{\tiny trap}}$ in (\ref{gtrap}), and we would like to focus on trajectories that don't become temporarily trapped (then $\Delta N_\perp$ can be non-analytic, Sec.~\ref{sec:numerics}), we require  $\mu>\mu_{an}$ from (\ref{muan}) in this section. Under these assumptions, the slowing down effect is recovered well by the approximations in Sec.~\ref{sec:slowingdown}. 

 Since we do not want the COBE normalization to be saturated by slow--roll contributions, that is by perturbations in $\varphi_\parallel$, we require \cite{Komatsu:2010fb}
\begin{eqnarray}\label{COBE}
\left(\frac{H_*}{2\pi}\right)^2\frac{1}{2\varepsilon_*}\ll \mathcal{P}_{\mbox{\tiny COBE}}\approx 2.41\times 10^{-9}\,,\label{mCOBE}
\end{eqnarray}  
or \begin{eqnarray}
m\ll m_{\mbox{\tiny COBE}}\equiv 6.2\times 10^{-6}\,.\label{mCOBE}
\end{eqnarray} 
This amounts to less fine tuning of the potential, since only an upper limit is imposed. On the other hand, the impact parameter needs to be just right to saturate the COBE bound, that is $x=\pi g\mu^2/v_p$ needs to solve
\begin{eqnarray}
xe^{-2x}=D \label{xfinetunedesp}\,,
\end{eqnarray}
where we used (\ref{deltaNslperp}) and defined 
\begin{eqnarray}
D&\equiv& \left(\mathcal{P}_{\mbox{\tiny COBE}}\left(\frac{2\pi}{H_*}\right)^2-\frac{1}{2\varepsilon_*}\right)\frac{1}{A^2C^2} \label{D}\\
&=& \mathcal{P}_{\mbox{\tiny COBE}}\left(\frac{2\pi}{H_*}\right)^2\frac{\Xi}{A^2C^2}\\
&=&\Xi\frac{2^63^4\pi^7}{g^6}\mathcal{P}_{\mbox{\tiny COBE}} \approx 1.6\times 10^7\frac{\Xi}{g^6}\mathcal{P}_{\mbox{\tiny COBE}}   \label{Dcobe}
\end{eqnarray}
with $A$ and $C$ from (\ref{ACx}). If slow--roll contributions are negligible $\Xi\rightarrow 1$ (the proper value for $\Xi$ is kept in all plots). For $g>g_{\mbox{\tiny COBE}}$, $D$ is smaller than the maximum value of $xe^{-2x}$ at $x_c=1/2$ (corresponding to $\mu=\mu_c$), that is $D<1/(2e)$; then equation (\ref{xfinetunedesp}) has the two real solutions
\begin{eqnarray}
x_i\equiv -\frac{1}{2}W_i(-2D)\,, \label{xm1}
\end{eqnarray}
for $i=0,-1$, where $W_i$ is the Lambert function. $i=0$ leads to an impact parameter smaller than $\mu_c$, and $i=-1$ to a bigger one; we focus on $i=-1$ primarily because impact parameters smaller than $\mu_c$ lead to a large negative $f_{NL}$ (see Fig.~\ref{pic:PandNLnumvsan}), that is observationally disfavoured\footnote{We also want to prevent a tuning of $\mu$ smaller than quantum fluctuations in any given Hubble time, $\delta\varphi_\perp^{QM} \sim H/(2\pi)$. Requiring $\mu_c/\delta\varphi_\perp^{QM}\gg 1$ amounts to $\varphi_p\sqrt{g}\ll 2/\sqrt{m}$ where $m<m_{\mbox{\tiny COBE}}$. \label{footnoteCOBE}} and define 
\begin{eqnarray}
\bar{x}\equiv x_{-1}=-\frac{1}{2}W_{-1}(-2D)\,. \label{barx}
\end{eqnarray}
This lower branch of the Lambert function is defined for arguments in the interval $[-1/e,0]$, but it has no simple series expansion at the boundaries (we are particularly interested in arguments close to the lower boundary of the interval). 

How well do we have to tune $x$ and thus the impact parameter to be within
$1\%$ of $\mathcal{P}_{\mbox{\tiny COBE}}$? From Eq.~(\ref{xfinetunedesp})
we find $xe^{-2x}=D\times(1\pm 10^{-2})$ leading to $x=\bar{x}\pm \delta
x$ with  \mbox{ $\delta x=|10^{-2}W_{-1}(-2D)/(2D(1+W_{-1}(-2D)))|\approx
0.08\sim \mathcal{O}(10^{-1})$}, where we performed a Taylor expansion
around $\bar{x}$ and took\footnote{In the limit $D\rightarrow 0$ the
tuning becomes more severe since $\delta x\rightarrow 0$ as well.} $D\sim
10^{-1}$ so that $\bar{x}\approx 1.3$ in the last step; this corresponds
to a tuning of about $6\%$ of $\bar{x}$, or about $3\%$ of
\begin{eqnarray}
\bar{\mu}\equiv \mu_c \sqrt{2\bar{x}}\,.\label{barmu}
\end{eqnarray}
We observe $\bar{\mu}\gtrsim \mu_c\gg\delta\varphi_{\perp}^{QM}$ as long
as $\varphi_p\sqrt{g}\ll 2/\sqrt{m}\gtrsim 2/\sqrt{m_{\mbox{\tiny
COBE}}}\sim 800$. For large $m$ close to $m_{\mbox{\tiny COBE}}$ and
$\varphi_p\sim 10$, the required tuning of the impact parameter $\delta
\bar{\mu}$ becomes uncomfortably close to $\delta\varphi_{\perp}^{QM}$ if
$g\sim 1$; hence, we require the inflaton mass to be at least a factor of
$5$ smaller than $m_{\mbox{\tiny COBE}}$ to be on the safe side.

If $m<m_{\mbox{\tiny COBE}}$, then $\mathcal{P}_{\mbox{\tiny COBE}}$ is
matched by a mix of slow--roll and slowing--down contributions.
In the limit $m/m_{\mbox{\tiny COBE}}\ll 1$
the slow--roll contributions become negligible, $\Xi\rightarrow 1$, and the scalar spectral index (\ref{ns-1}) becomes
\begin{eqnarray}
n_s-1\approx 0
\end{eqnarray}
to first order in slow--roll parameters, since $\eta_{\perp}^*=\varepsilon^*$. An $n_s$ this close to one is observationally disfavoured \cite{Komatsu:2010fb} at the two sigma level\footnote{For more discussion of the spectral index in a related scenario see \cite{Byrnes:2005th}. At second order in slow roll, $n_{s}-1=-10\varepsilon^2/3$ \cite{Byrnes:2006fr}.}. However, a scale dependence in line with observations can easily be introduced by considering more general polynomials in (\ref{sqpot}) with $\varepsilon^*\neq \eta^*_{\perp}$ (we leave this to future studies). The tensor--to-scalar ratio in (\ref{scalartotensor}) approaches zero in the limit $\Xi\rightarrow 1$ regardless of the potential. Thus, a detection of gravitational waves of order $r\sim 0.1$ would rule out a grazing ESP encounter as the dominant contribution of the power-spectrum. 

\begin{figure}[tb]
\begin{center}
\includegraphics[scale=0.6,angle=0]{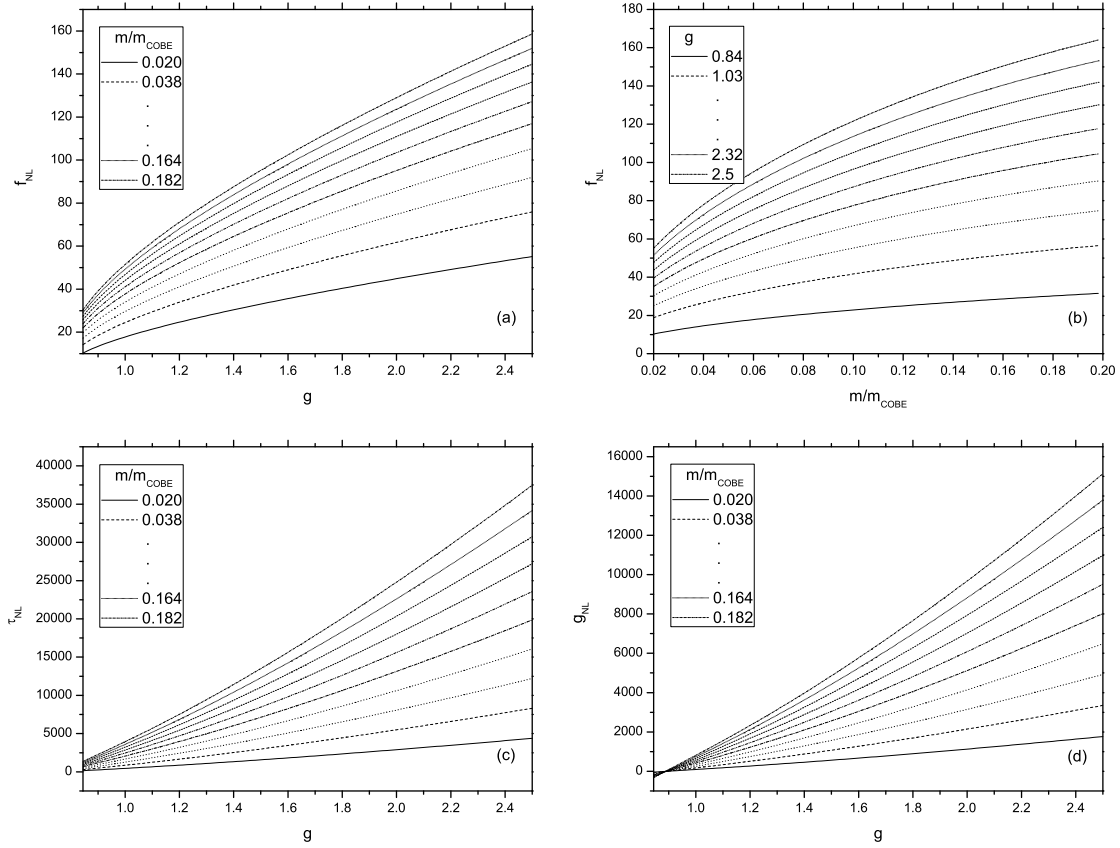}
   \caption{\label{pic:NLftESP} The non-linearity parameters $f_{NL}$, $\tau_{NL}$ and $g_{NL}$ in (\ref{fNLsl})-(\ref{gNLsl}) from the slowing down effect, evaluated at $\mu=\bar{\mu}$ in (\ref{barmu}) for which the COBE normalization is satisfied; slow--roll contributions are negligible for the power-spectrum, since $m\ll m_{\mbox{\tiny COBE}}=6.2\times 10^{-6}$; we vary $0.02<m/m_{\mbox{\tiny COBE}}<0.2$ and $1.1 \times g_{\mbox{\tiny COBE}}<g<2.5$. For masses not too small compared to $m_{\mbox{\tiny COBE}}$, non-Gaussianities are observable and keep increasing with $g$. Results are indicative even for $g>g_{\mbox{\tiny trap}}$ as long as $\bar{\mu}>\mu_{an}$ (satisfied above), but since $g_{\mbox{\tiny trap}}>g_{\mbox{\tiny max}}$ in (\ref{gmax}), the sudden particle production approximation ceases to be valid and effects may be overestimated.}
   \end{center}
\end{figure}

As discussed in appendix \ref{app:nonadd},  the impact parameter needs to be smaller than $\mu_{\mbox{\tiny max}}=\sqrt{v_p/g}$,
to guarantee a large non-adiabaticity parameter and allow for particle production to be treated as instantaneous. Since $\bar{\mu}/\mu_{\mbox{\tiny max}}=1/\sqrt{2\pi}$, we see that $\bar{x}<\pi$ has to hold. Solving $\bar{x}(D)=\pi$ using (\ref{barx}), leads to $D_{\mbox{\tiny min}}\approx 0.0059$, which in turn leads to an upper bound on the coupling $g$ via (\ref{Dcobe}) of
\begin{eqnarray}
g<g_{\mbox{\tiny max}}\equiv \left(\Xi\frac{2^63^4\pi^7}{D_{\mbox{\tiny min}}}\mathcal{P}_{\mbox{\tiny COBE}}\right)^{1/6}\approx 1.4\,,\label{gmax}
\end{eqnarray}  
where we used $\Xi\approx 1$ in the last step. Therefore results for coupling constants above $g_{{\mbox{\tiny max}}}$ and $\mu=\bar{\mu}$, i.e.~in Fig.~\ref{pic:NLftESP}, should be taken with caution, since our approximations start to become unreliable\footnote{One could employ the iterative analytic approximation scheme in Appendix C of \cite{Kofman:2004yc} (an expansion for small non-adiabaticity parameters) if $g\gg g_{\mbox{\tiny max}}$ and
 $\mu=\bar{\mu}$, but since we expect non-Gaussianities to grow uncomfortably large, we refrain from applying it here.}.
 
To briefly summarize, given a value of the coupling in the interval $ 1.4 \approx g_{\mbox{\tiny max}}>g>g_{\mbox{\tiny COBE}}\approx 0.77$ and $m\sim m_{\mbox{\tiny COBE}}/\mbox{few}$, the impact parameter  needs to be $\mu=\bar{\mu}$ within a few percent in order for the power-spectrum to be dominated by fluctuations in $\varphi_{\perp}$ while satisfying the COBE normalization, with $n_s-1\approx 0$ (different potentials can provide a red tilt) and unobservable small tensor--to--scalar ratio $r$. Next to the suppression of primordial gravitational waves, how can we tell this mechanism apart from a simple slow roll model?

Since the above mechanism is contingent on a relatively strong dependence of the number of efoldings onto the isocurvature direction $\varphi_{\perp}$, one should expect large non-Gaussianities \footnote{Since we use a quadratic potential and equal masses for the two fields, the scale dependence of $f_{NL}$, see forward to Sec.~\ref{sec:modulatedtrapping}, is negligible. If a different power-law were used, a scale dependence of the non-linearity parameters would follow.}. This is indeed readily checked, by plugging $\bar{x}$ of (\ref{xm1}) into the expressions for the non-linearity parameters (\ref{fNLsl})-(\ref{gNLsl}). A plot of these parameters over $g>1.1\times g_{\mbox{\tiny COBE}}$ is given in Fig.~\ref{pic:NLftESP}, where we chose $\varphi_p\approx 11$, so that the ESP is encountered around $30$ efoldings before the end of inflation; we varied the inflaton mass in the range $0.02<m/m_{\mbox{\tiny COBE}}<0.2$.

First, we observe that $f_{NL}$ is generically observable ($\mbox{few}<f_{NL}<100$) for $g_{\mbox{\tiny COBE}}<g<g_{\mbox{\tiny max}}$ and $m=m_{\mbox{\tiny COBE}}/\mbox{few}$. $f_{NL}$  increases monotonically with increasing $g$ and decreases as the inflaton mass is decreased. As a consequence, for low inflaton masses, a relatively broad range of coupling constants permits observable non-Gaussianities that are not yet ruled out, but results become untrustworthy above $g>g_{\mbox{\tiny max}}$. Values consistent with current observations are located around $g\sim 1$, the most natural value for the coupling constant, so no extra tuning is needed. The non-linearity parameters of the trispectrum are naturally of order $f_{NL}^2$ (Fig.~\ref{pic:NLftESP}): 
\begin{eqnarray}
\frac{\tau_{NL}}{f_{NL}^2}\bigg|^{sl}_{x=\bar{x}}
&=&\left(\frac{6}{5}\right)^2\frac{1}{\Xi}\approx \left(\frac{6}{5}\right)^2\,,\label{consistency1}\\
\frac{g_{NL}}{f_{NL}^2}\bigg|^{sl}_{x=\bar{x}}&=&\frac{2}{3}\frac{\left(\bar{x}-\frac{3}{2}\right)\bar{x}}{\left(\bar{x}-\frac{1}{2}\right)^2}\frac{1}{\Xi}\approx \frac{2}{3}\frac{\left(\bar{x}-\frac{3}{2}\right)\bar{x}}{\left(\bar{x}-\frac{1}{2}\right)^2}\,.\label{consistency2}
\end{eqnarray}
$\tau_{NL}/f_{NL}^2$ is independent of $g$ and only weakly dependent on $m$ via $H_*$ in $\Xi$, while $g_{NL}/f_{NL}^2$ also depends on $g$ via $\bar{x}$. It is clear that measuring both the bi- and trispectrum would help to distinguish this model from other scenarios \cite{Suyama:2007bg,Suyama:2010uj} (see \cite{Suyama:2011qi} for relationships between higher--order parameters). See the introduction for constraints on these parameters.

We put the ESP encounter around $N_p\approx 30$ efoldings before the end of inflation,  well after observable scales left the horizon, for a good reason: if the ESP encounter were to take place close to or within the observational window, back-scattering of $\chi$ particles onto the inflaton condensate would lead to an additional bump-like contribution to the power-spectrum and non-Gaussianities via IR cascading \cite{Barnaby:2009mc,Barnaby:2009dd,Barnaby:2010ke,Barnaby:2010sq}. This unobserved contribution would dominate the power-spectrum (and non-Gaussianities) for $g\gtrsim 0.1$ if $\mu$ 
is small\footnote{The bound $g\lesssim 0.1$  is derived in the $\mu=0$ limit. As $\mu$ grows, the exponential suppression in (\ref{bump}) suppresses the amplitude so that larger $A_{b}\propto g^{15/4}$ and thus $g$ are possible. In our case, $\pi g\bar{\mu}^2/v_p=\bar{x}=-W_{-1}(-2D)/2$ is close to one so that the bound is not raised much (for example, $\bar{x}\sim 1.3$ for $D\sim 10^{-1}$ so that $g\gtrsim 0.14$ instead of $0.1$).}, well before the ESP encounter starts to matter. But as long as the ESP is grazed later, the bump like feature is on smaller scales which are not tightly constrained by observations (but might be in the future). Choosing $N_p$ different from $30$ leads to changes of order $\mathcal{O}(1)$ in the constants $A,B,C$ in (\ref{ACx}) and (\ref{B}), but $\bar{x}$  in (\ref{barx}) is insensitive, since $D$ in (\ref{D}) depends only weakly on $\varphi_p$. As a consequence, the non-linearity parameters in (\ref{fNLsl})-(\ref{gNLsl}) are simply rescaled in the $\Xi\rightarrow 1$ limit ($f_{NL}\propto A\propto 1/\phi_p\propto 1/\sqrt{N_p}$ and $\tau_{NL},g_{NL}\propto A^2\propto 1/\phi_p^2\propto 1/N_p$) and the consistency relations in (\ref{consistency1}) and (\ref{consistency2}) remain unaffected. 

The observation of a single bump caused by backscattering in the vicinity of $k_{b}=\sqrt{gv_p}\exp(N_{b})$ where $N_{b}=N-N_p$,  approximately\footnote{This approximation does not retain oscillations in the tail of the exponential \cite{Barnaby:2009mc,Barnaby:2009dd}.} given by 
\begin{eqnarray}
\mathcal{P}_{b}\approx A_{b}\left(\frac{\pi e}{3}\right)^{3/2}\frac{k^3}{k_{b}^3}e^{-\frac{\pi}{2}\frac{k^2}{k_{b}^2}}e^{-\pi\frac{g\bar{\mu}^2}{v_p}}\,,\label{bump}
\end{eqnarray}
with $g\sim 1$ and  $A_{b}\approx 10^{-6}g^{15/4}$ \cite{Barnaby:2009mc,Barnaby:2009dd}, is a smoking gun of a grazing ESP encounter. Note that $\pi g\bar{\mu}^2/v_p=\bar{x}=-W_{-1}(-2D)/2$. The location of this feature determines when the ESP has been encountered, and the height offers a consistency check of the model, since no free parameters are left. This bump is accompanied by large, potentially detectable, non-Gaussianities \cite{Barnaby:2010ke,Barnaby:2010sq} in the vicinity of $k_b$. 

It should be noted that all of the above arguments remain valid in higher dimensional field spaces and more general polynomial potentials (as long as all fields have the same potential); then the two fields that we consider above  parametrize the plane spanned by the unperturbed trajectory and the ESP, while contributions from additional perpendicular directions are ignored. 

If the potential is more complicated, one may still compute additional contributions to correlation functions in a similar manner, with the appropriate changes in i.e.~the relationship between $\mu_p$ and $\varphi_\perp^*$ in (\ref{muandphistar}). If the trajectory goes through an effective single-field stage after the observational window is passed but before the ESP is encountered, all sensitivity to the initial field values are erased and no additional super-horizon fluctuations are imprinted onto cosmological fluctuations when the ESP is passed. By single-field stage we mean a regime where all trajectories are funnelled through a valley in the potential such that all but one direction have a heavy mass. 

So far, we just assumed that an ESP is closely grazed, but how likely is such a single encounter? 

\subsection{Several ESP encounters and connection to trapped inflation \label{Sec:trappedinflation}}

If ESPs are dense we expect more than one close encounter during the last sixty efoldings of inflation. If $\chi$-particles dilute sufficiently in between encounters and none of them involves actual trapping, the contributions to the power-spectrum and higher--order correlation functions are additive, and the results of the previous section can be used directly. However, if encounters start to overlap, that is if one or more ESPs are passed per Hubble time, their combined backreaction can lead to a reduction of the speed throughout inflation, enabling steeper potentials to be used. This phenomenon is sometimes referred to as trapped inflation \cite{Kofman:2004yc,Green:2009ds,Silverstein:2008sg,Battefeld:2010sw}, which has been investigated in one dimensional \cite{Green:2009ds,Silverstein:2008sg} and higher dimensional \cite{Battefeld:2010sw} field spaces.

For simplicity, let us assume that ESPs are distributed evenly over field space with some average inter-ESP distance $y$. In trapped inflation, the sensitivity of the inflationary trajectory to individual ESP encounters is greatly reduced: a change in $\varphi_\perp^*$ lowers the distance to about one half of the relevant ESPs, while the distance to the other half is raised. Since the contributions to the number of efoldings $N$ from each ESP encounter are additive, the dependence of $N$ on $\varphi_\perp^*$ vanishes in the limit of $y\rightarrow 0$. Thus, the effects computed in the previous sections are diminished for $y$ smaller than some critical inter-ESP separation $y_{crit}$. 

Let us estimate this critical inter-ESP distance for a field-space of arbitrary dimensionality $D$. We assume the same quadratic potential for all inflatons, so that it has spherical symmetry. In the absence of ESPs, any trajectory from $|\vec{\varphi}|=\varphi_{*}\sim 15$ to  $|\vec{\varphi}|=\varphi_{end}\sim 1$ yields the same inflationary dynamics. Thus, the total volume in which a slow--roll trajectory could lie is $V_{tot}=(\varphi_{*}^D-\varphi_{end}^D)V_D$ where $V_D=\pi^{D/2}/\Gamma(D/2+1)$ is the volume of a unit sphere in $D$ dimensions. On the other hand, a grazing ESP encounter would only lead to observationally relevant contributions to observables if the ESP is within a distance of $\mu_c\sim 10^{-3}$ to a given inflationary trajectory. If ESPs are randomly spread over field space, we can compute the probability that a single ESP in $V_{tot}$ is within reach of the trajectory to
\begin{eqnarray}
p_{\mbox{\tiny ESP}}=\frac{V_{D-1}(\varphi_{*}-\varphi_{end})\mu_c^{D-1}}{V_{D}(\varphi_{*}^D-\varphi_{end}^D)}\,.
\end{eqnarray}
Thus, if $N_{\mbox{\tiny ESP}}$ ESPs are spread throughout $V_{tot}$, the probability that none of them is closely grazed becomes $(1-p_{\mbox{\tiny ESP}})^{N_{\mbox{\tiny ESP}}}$. Demanding that the latter is less than $1/2$ leads to the minimal number of ESPs in $V_{tot}$ above which ESP encounters become likely,
\begin{eqnarray}
N_{\mbox{\tiny ESP}}^{min}=\ln\left(p_{\mbox{\tiny ESP}}-\frac{1}{2}\right)\,.
\end{eqnarray}
The corresponding average inter-ESP distance is
\begin{eqnarray}
y_{crit}=\left(\frac{V_D(\varphi_{*}^D-\varphi_{end}^D)}{N_{\mbox{\tiny ESP}}^{min}}\right)^{1/D}\,.
\end{eqnarray}
Note that $y_{crit}\rightarrow 0$ in the limit of $D\rightarrow \infty$. For $D=2,10,100$ the minimal number of ESPs is $N_{\mbox{\tiny ESP}}^{min}\approx 1.7\times 10^4, 10^{38}, 10^{413}$ with corresponding average inter-ESP distances of $y_{crit}\approx 0.20, 0.0030, 0.00045$. Thus, for low $D$, ESPs do not need to be particularly dense in order for a grazing ESP encounter to be likely.

However, with growing dimensionality $D$, it becomes increasingly unlikely that only a few ESP encounters take place: either $y>y_{crit}$ and no ESP is encountered, or $y$ is (slightly) below $y_{crit}$ and many ESPs are grazed. Thus, in the large $D$ limit we either expect undisturbed slow--roll inflaton, or trapped inflation at a terminal velocity \cite{Battefeld:2010sw}.

If trapped inflation takes place, the dominant correction, next to the ones stemming from a reduced speed, originate from backscattering of $\chi$-particles onto the inflaton condensate, which superpose to yield yet another nearly scale invariant contribution to the power-spectrum, accompanied by non-Gaussianities \cite{Barnaby:2009mc,Barnaby:2009dd,Barnaby:2010ke,Barnaby:2010sq}. This superposition of bumps in the power-spectrum has been investigated in \cite{Barnaby:2009dd} for trapped inflation in one dimension. The corresponding computation in higher dimensional field spaces for the power-spectrum and the three-point function is in progress \cite{inprep}\footnote{Preliminary results show that the additional contribution to the power-spectrum
acquires a blue tilt; hence, it cannot be the dominant contribution, but non-Gaussianities might still be observable.}, but goes beyond the scope of the present article. 

To summarize, for inter-ESP distances of order $y_{crit}$ and low $D$, several ESP encounters are likely, without leading to trapped inflation, and the results of this article are valid, namely, the dominant contribution to the power-spectrum can originate from individual ESP-encounters, naturally leading to observably large non-Gaussianities.   

\section{Observational signatures from modulated trapping \label{sec:modulatedtrapping}}

We now consider the model of modulated trapping introduced in \cite{Langlois:2009jp}. In this scenario, there is only one inflaton field, so that the inflaton trajectory is single-dimensional, but there exists another field, the modulaton $\sigma$, which  affects the strength of the trapping effect. As we will see below, the r\^ole  of the modulaton on the primordial curvature perturbation is quite analogous to that of the perpendicular inflaton perturbation $\delta\varphi_\perp$ discussed in the previous sections. 

More explicitly, the interaction between the inflaton $\varphi$ and a bosonic field $\chi$ is of the form
\bea\label{LS} {\mathcal L}=-\frac12(m(\sigma)-\lambda(\sigma)\varphi)^2\chi^2, \eea
leading to  a trapping event at time $t_p$, with corresponding field value
\bea \varphi_p=\frac{m(\sigma)}{\lambda(\sigma)}\,. \eea
Both the time and the strength of the trapping event 
 depend on $\sigma$, the modulaton field, which we assume to be light compared to the Hubble parameter during inflation. 

However, in contrast to \cite{Langlois:2009jp} we will not neglect its mass, and so its energy density after the trapping event should be taken into account. We will see that the modulatons potential 
in general also affects the spectral index and the scale dependence of $f_{NL}$.  If the modulaton field decays after the end of inflation, it acts as a curvaton field. In this section we assume that the energy density due to the modulaton during inflation is negligible (in order to be able to neglect its effect after inflation), so that
\bea 
U(\varphi,\sigma)=V(\varphi)+W(\sigma)\simeq V(\varphi)\simeq 3 H^2,\qquad V_{,\varphi}\gg W_{,\sigma}\;\Rightarrow\; \varepsilon\equiv\varepsilon_{\varphi}\simeq\varepsilon_H\gg\varepsilon_{\sigma}\,. 
\eea
However, the above relations leave the relative values of the second derivatives undetermined, so it is possible to have $1\gg|\eta_{\sigma}|>|\eta_{\varphi}|$. 

We can write the primordial curvature perturbation as
\bea \zeta(\bk)=\zeta^G_{\varphi}(\bk)+\zeta^G_{\sigma}(\bk)+ \frac35f_{\sigma}(k)\left(\zeta^G_{\sigma}\star\zeta^G_{\sigma}\right)(\bk) + \frac{9}{25}g_{\sigma}(k)\left(\zeta^G_{\sigma}\star\zeta^G_{\sigma}\star\zeta^G_{\sigma}\right)(\bk)\,, \eea
where $\zeta^G_{\varphi}$ and $\zeta^G_{\sigma}$ are Gaussian, and it is a good approximation to treat the contribution from the fields as uncorrelated \cite{Byrnes:2006fr}. The quantity $f_{\sigma}$, which is $\fnl$ in the limit that we neglect the inflaton field's perturbations (and similarly for $g_{NL}$), is given by
\bea \frac65f_{\sigma}=\frac{\Delta N_{,\sigma\sigma}}{(\Delta N_{,\sigma})^2}, \qquad \frac{36}{25}g_{\sigma}=\frac{\Delta N_{,\sigma\sigma\sigma}}{(\Delta N_{,\sigma})^3}\,, \eea 
where
\bea \Delta N=\frac{ \lambda^{5/2}}{18\pi^3}\frac{M_P^{1/2}}{H_{p}^{1/2}} (2\varepsilon_{p})^{1/4}=\Delta N(\sigma_{p})\,. \eea
The power-spectrum is given by
\bea P_{\zeta}=P_{\zeta_{\varphi}}+P_{\zeta_{\sigma}}=P_{\zeta_{\varphi}} (1-\Xi)^{-1}, \eea
where we have defined $\Xi(k)$ analogously to (\ref{Xi}) and the spectral index satisfies, (see (\ref{ns-1})),
\bea\label{spectralindex}
n_s-1=-(6-4\Xi)\varepsilon+2(1-\Xi)\eta_{\varphi}+2\Xi\eta_{\sigma}\,.   
\eea
Then the non-Gaussianity and its scale dependence is given by \cite{Byrnes:2009pe,Byrnes:2010ft}
\bea\label{fNL} 
\fnl&=&\Xi^2(k)f_{\sigma}(k)\,, \\ 
\nfnl\equiv \frac{\partial\ln|\fnl|}{\partial\ln k} &=&4(1-\Xi)(2\varepsilon+\eta_{\sigma}-\eta_{\varphi})+\frac{\partial\ln|f_{\sigma}|}{\partial\ln k} \nonumber
  \\ &=&4(1-\Xi)(2\varepsilon+\eta_{\sigma}-\eta_{\varphi}) +\frac{\Delta N_{,\sigma}}{\Delta N_{,\sigma\sigma}}\frac{W_{,\sigma\sigma\sigma}}{V}M_P^2\,, \label{nfNL} \\ \tau_{NL}&=&\frac{1}{\Xi}\left(\frac65 f_{NL}\right)^2\,, \\
   g_{NL}&=&\Xi^3g_{\sigma}\,. \label{modtrapping:gNL}\eea
Note that $\Xi$ is an observable if we can observe the bi- and trispectrum \cite{Smidt:2010ra}. All terms, except those involving $\Delta N$, in the above equations should be evaluated at horizon crossing. 
$\nfnl$ is the sum of two independent parts: the first 
one is due to the presence of linear
contributions from two scalar fields, as was discussed in Sec 2 of \cite{Byrnes:2009pe}, see in particular the mixed inflaton-curvaton scenario; this part vanishes if one neglects the perturbations due to the inflaton field ($\Xi=1$). The second part is due to non-linear interactions of the modulon field's perturbations, which vanish if the modulaton field has a quadratic potential, since the modulaton field perturbations obey a linear equation of motion (see Sec.~3 of \cite{Byrnes:2009pe}, in particular the interacting curvaton scenario). Observational prospects for $\nfnl$ are considered in \cite{Sefusatti:2009xu,Shandera:2010ei}.

In Sec.~\ref{sec:grazingESP}, we considered two fields with equal mass and a quadratic potential (implying that all three first order slow--roll parameters are equal), so the scale dependence of $f_{NL}$ is especially simple in this case, satisfying two independent consistency relations with the tensor--to--scalar ratio and spectral index
\bea r=2n_{f_{NL}}=-4(n_s-1)=16(1-\Xi)\varepsilon_*\,. \label{consitencyrelrunning}\eea
In the limit that perturbations come from the (initially) isocurvature field and $\Xi\rightarrow 1$ all observables, except for the three non-linearity parameters and the amplitude of the power-spectrum, are zero.

\subsection{Special cases \label{sec:modtrapspecialcases}}
In \cite{Langlois:2009jp} a special case was considered for which $W(\sigma)$ and all of its derivatives are negligible, with the consequence that
\bea 
n_s-1&=&-(6-4\Xi)\varepsilon+2(1-\Xi)\eta_{\varphi}\,, \\ \nfnl&=&4(1-\Xi)(2\varepsilon-\eta_{\varphi})\,. \eea
If modulated trapping provides the dominant contribution to the power-spectrum, $\fnl$ is scale independent. Alternatively, if $\Xi$ is not too close to unity and $\varepsilon\ll|\eta_{\varphi\varphi}|$, a non-trivial consistency relation holds:
\bea \nfnl=-2(n_s-1)\,. \eea

In the special case of a {\it modulaton dependent coupling}, that is
\bea\label{dependentcoupling} \lambda=\frac{\sigma}{M}\,,\qquad m=\tilde{g}\sigma \,, \eea
$\varphi_p$ is independent of $\sigma$. Then $\Delta N\propto\sigma_p^{5/2}$ and it follows that $\Delta N_{,\sigma}/\Delta N_{,\sigma\sigma}=2\sigma_p/3$. Assuming $W(\sigma)\propto\sigma^q$ we find
\bea  \nfnl\equiv \frac{\partial\ln|\fnl|}{\partial\ln k} =4(1-\Xi)(2\varepsilon+\eta_{\sigma}-\eta_{\varphi}) +\frac23(q-2)\eta_{\sigma}\,.
\eea

\subsection{The modulaton as a curvaton \label{sec:modulatonascurvaton}}
After the trapping event the modulatons energy density needs to be taken into account. If $\sigma$ decays after the inflaton does, it can act as a curvaton, adding a third term to the curvature perturbation
\bea \zeta=\zetai+\zetat+\zetac. \eea
Since $\zetat\propto\zetac\propto\delta\sigma_*$ the two latter terms are 
correlated, while $\zetai\propto\delta\varphi_*$ is uncorrelated to them. Unlike $\zetac$, $\zetat$ usually does not lead to any isocurvature perturbations since any particles created during the ESP encounter, 
and their decay products will be quickly diluted by the subsequent inflation, unless the encounter is close to the end of inflation. 
Due to the correlation properties of the three terms, we extend the definition of $\Xi$ to
\bea \Xi=\frac{P_{\zetat}+P_{\zetac}}{P_{\zeta}}, \eea
where
\bea P_{\zeta}=P_{\zetai}+P_{\zetat}+P_{\zetac}. \eea
Then all of the formula's given in (\ref{spectralindex})--(\ref{modtrapping:gNL}) for the spectral index and the non-linearity parameters remain valid.

We would like to comment on a related idea: in the case of modulated reheating, the field which modulates the efficiency of reheating can also act as a curvaton \cite{Suyama:2010uj}.  
One could explicitly calculate both the adiabatic and isocurvature perturbations arising from this scenario, at the level of the power- and bispectrum using the formalism developed in \cite{Langlois:2011zz}, and at the level of the trispectrum using the formalism in \cite{Langlois:2010fe} which includes perturbations up to third order. We leave this project to future research.

\section{Relating a grazing ESP encounter and modulated trapping \label{sec:grazESPandModulatedTrapping}}
We have investigated two possibilities of how an ESP encounter can modulate perturbations in the presence of a second field: firstly, by putting the ESP in a two dimensional field space so that small fluctuations in the field perpendicular to the trajectory modulate the impact parameter and thus particle production, Sec.~\ref{sec:grazingESP}. Secondly, by traversing the ESP head on, but including a dependence of the coupling to $\chi$ and/or its bare mass on a second scalar field, Sec.~\ref{sec:modulatedtrapping}. How are these two scenarios related?
  
The modulated trapping model is more general in the sense that the strength of the coupling and the position of the ESP is allowed to depend on the isocurvature field. However, it is significantly more restricted in the sense that a straight trajectory is assumed throughout inflation that traverses the ESP, i.e.~the $\sigma$ field is not treated as an inflationary direction; consequently, no exponential suppression due to a non-zero impact parameter is possible (see e.g.~the last term in (\ref{DeltaNsl})). 
Hence neither scenario is a subset of the other, but they are similar in spirit and in certain cases may become effectively equivalent.

Let us consider a modulaton-dependent coupling as in (\ref{dependentcoupling}) (the only explicit model considered in \cite{Langlois:2009jp} 
which can generate $\Xi\simeq 1$ with only one species of $\chi$ particles), but also allow the model to be two-dimensional, that is we identify the perpendicular direction in Sec.~\ref{sec:grazingESP} with the modulaton $\sigma$. Then the change in the number of efoldings due to decreasing speed in field space in (\ref{DeltaNsl}) becomes
\bea\label{mdc:N} \Delta N_{sl}\simeq \frac{\sigma^{5/2}}{M^{5/2}} \frac{10^{-4}}{\sqrt{m} \varphi_p } e^{-3\sigma^3/(M\sqrt{m})} \simeq \frac{\sigma^{5/2}}{M^{5/2}} 10^{-2} e^{-3\sigma^3/(M\sqrt{m})}, \eea
where we have chosen typical values $\varphi_p\sim10$ and $m\sim10^{-6}$ after the second $\simeq$ sign in order to simplify the analysis. This leads to
\bea \Delta N_{sl,\sigma}\propto \frac52\frac{\sigma^{3/2}}{M^{5/2}} - 9 \frac{\sigma^{9/2}}{\sqrt{m}M^{7/2}}\,. \eea
We see that the contribution from the exponential term dominates unless $\sigma^3\lesssim10^{-3}M$, which is about the same condition for being able 
to neglect the exponential suppression in (\ref{mdc:N}). Then in order to satisfy the COBE normalisation ${\cal P}_{\zeta}\sim 10^{-10}$ (see (\ref{COBE})), we need to satisfy $\sigma^3 H_*^2 10^{-4}/M^5=10^{-10}$; hence, for this model to work in the regime of \cite{Langlois:2009jp} we require $M\lesssim10^{-2}$ and $|\sigma|\lesssim 10^{-2}$. These conditions are  
consistent with those given in eq.~(43) of \cite{Langlois:2009jp}, but here we have also checked whether or not one can neglect the exponential suppression term. We see that there exists a consistent regime, but within a fine tuned range of initial  
values for $\sigma$.

\subsection{Modulated trapping in a 2-dimensional field space}
The effect of the exponential suppression and fine tuning of the initial conditions is potentially reduced if we consider an Extra Species Locus (ESL) instead of an ESP. Two simple choices come to mind: a straight line ESL along $\varphi=\varphi_p$, so that every trajectory passes through the ESP during inflation, or a compact ESL i.e.~a circular one centred at the origin. Both choices are feasibly in moduli spaces.
since they simply indicate that the ESL is independent of a field (the first choice) or an angular direction. We choose to focus on the second case only, without any theoretical bias either way. To make the two-dimensional field space explicit, we replace $\sigma$ by $\varphi_2$ from here on. Thus, we wish to investigate an interaction Lagrangian of the form
\bea\label{modESL} {\cal L}=-\frac12 g^2 \left(|\stackrel{\rightarrow}{\varphi}|- |\stackrel{\rightarrow}{\varphi_p}|\right)^2 \chi^2 \,, \eea
with a modulated coupling, $g^2=\varphi_2^2/M^2$. Then all trajectories become nearly straight lines and the exponential suppression term can be neglected for any trajectory. However, a simple calculation shows that approximately the same fine-tuning on the initial condition for $\varphi_2$ is needed in order to have the correct amplitude of the power-spectrum if these perturbations dominate over the usual slow--roll contribution, i.e.~$|\varphi_2|\lesssim10^{-2}$ is still required. 

If we had considered a constant coupling $g$, then all trajectories would be equally delayed by the ESL encounter and there would be
no effect on the perturbation spectrum (in the case of a rotationally symmetric potential, which we are considering here). However, in general even an ESL with a constant coupling will generate perturbations provided that it breaks some symmetry between trajectories with an isocurvature perturbation between them. For example, an elliptical ESL centred at the origin, or a circular ESL not centred at the origin, will both generate a contribution to the curvature perturbation at the time of particle production. Such a model would be similar to an inhomogeneous end of inflation \cite{Lyth:2005qk,Salem:2005nd,Sasaki:2008uc,Naruko:2008sq,Byrnes:2008zy,Huang:2009vk}, which typically requires a very strongly elliptical surface on which inflation ends as well as a fine tuning of initial conditions \cite{Alabidi:2006wa}.

Why do we require this tuning, even when we are justified in neglecting the exponential suppression in (\ref{mdc:N})? A discussion of this point is given in the conclusion of \cite{Byrnes:2008zz}. In order for the perturbations associated with the trapping ($\varphi_2$ direction) to dominate over those generated by the inflaton in the usual way ($\varphi_1$ direction), we require that
\bea\label{isoc-inequality} \left(\frac{\partial N}{\partial \varphi_2}\right)^2 \gg \left(\frac{\partial N}{\partial \varphi_1}\right)^2\sim \frac{1}{\varepsilon_*} \gtrsim 10^2\,. \eea
The size of the isocurvature perturbations in each model is given by
$\delta\varphi_2={\cal O}(H)={\cal O}(m_1)$, independently of the background trajectory. Thus, these small perturbations correspond to a significant perturbation in the value of $\varphi_2$ only if the background trajectory lies close to $\varphi_2=0$, i.e.~$\delta\varphi_2/\varphi_2$ is non-negligible only for small $\varphi_2$. This requirement is needed to satisfy the inequality (\ref{isoc-inequality}) for all of the models that we have considered so far.
However, if $\varphi_2$ is too small, i.e. $\varphi_2={\cal O} (\delta\varphi_2)$, then non-Gaussianities become large and the model is ruled out by observations \footnote{In Sec.~\ref{sec:tunedESP} the axis are chosen such that the ESP lies on the $\varphi_1$-axis. The requirement that the value $\varphi_2$ is small is then identical to the requirement that the impact parameter $\mu$ to the ESP is small in Sec.~\ref{sec:tunedESP}. Further, $\varphi_2>{\cal O} (\delta\varphi_2)$ was written as $\varphi_\perp^*>\delta \varphi^{QM}$ which we imposed throughout Sec.~\ref{sec:grazingESP} and Sec.~\ref{sec:tunedESP}.}.

\section{Conclusions}

The possibility that the effective mass of a field coupled to the inflaton
becomes small, leading to particle production, has been well studied
in the case that this occurs either while modes of the scale observed
today cross the horizon (a window of 5--10 efoldings) or shortly after
inflation ends, i.e.~preheating. However, there is a long period in between
these two times (circa 50 efoldings) during which particle production may
occur, and this case has been much less studied. Although we cannot
directly observe any bump/feature this may cause on the power-spectrum on
these scales, simply because we are unable to observationally probe these
scales, we may see an effect on the usual CMB scales.

In the simplest case, with only one inflaton field, there is no affect onto
observables caused by such particle production, but if the strength or
position of the ESP depends on a second field, then this conclusion does
not hold, as was studied by Langlois and Sorbo \cite{Langlois:2009jp}.
Here we focussed instead on multiple--field inflation. For
most initial conditions, the trajectory in field space will not go through
the ESP, but graze it, still leading to particle production and backreaction in the vicinity of the ESP. 
The dominant effect is usually
the temporary slow down of the inflaton, as part of
its kinetic energy is
diverted to generating particles, 
which slow down the fields further via their backreaction; in addition
the trajectory bends towards
the ESP due to the attraction of the generated particles, whose mass grow
as the inflaton moves further away. The expansion 
of the universe
and associated
dilution
of these particles ensures that the effect only lasts a few efoldings
before a slow--roll attractor trajectory is again reached.

In multi-field scenarios the minimum distance between trajectories and the
ESP varies
due to the presence of isocurvature perturbations (those
perpendicular to the velocity  
along the inflaton trajectory). The closer to
the ESP that a trajectory comes, the greater the particle production and
the larger the consequent slow-down of the trajectory, hence the
isocurvature perturbation is partially converted to the adiabatic
perturbation on all scales which have already crossed the horizon. In
order for this encounter to significantly affect the trajectory, it must
pass close enough to be reasonably strongly distracted by the ``beauty" of
the ESP. For the models we have studied, an impact parameter of order
$10^{-3}M_P$ is required, which would represent a fine tuning of the
initial conditions if only one ESP exists, but for an average ESP spacing
of order $0.3 M_P$ at least one such encounter becomes likely. An order
of unity coupling constant between the inflaton and the produced particles
is the most interesting case, so no tuning of this parameter is required.
The case of a very dense set of ESP's, leading to multiple trapping events
per efolding, was previously studied by D.~and T.~Battefeld in
\cite{Battefeld:2010sw}.

What are the most interesting consequences of such a grazing ESP encounter?
Since this effect is asymmetric, most strongly affecting trajectories
which pass closest to the ESP and most weakly those furthest away, this
will generate a skewness in the probability distribution of trajectories,
which is parametrised by $f_{NL}$ (the amplitude of the bispectrum). The
trispectrum is also affected, and the ratio $\tau_{NL}/ (6
f_{NL}/5)^2\geq1$ provides a measure on which proportion of the observed
linear perturbation where generated by the ESP encounter, with the
inequality being saturated in the case that the encounter gives the dominant effect and the usual slow--roll perturbations are negligible. Since we have
studied a quadratic potential with equal masses the scale dependence of
the power spectrum and non-linearity parameters are suppressed, but for a
more complex potential they would typically be a larger.

We have also considered the possibility of a one dimensional enhanced
symmetry locus, which is a natural generalisation of the modulated
trapping model to multiple fields. In this case every trajectory may pass
through a point where the coupled field becomes massless, removing the
need to fine-tune initial conditions, but one needs a strong difference in
the strength of the particle production event between trajectories in
order generate an observably large signature.

In all cases we find observably large non-Gaussianities on CMB scales given that the power-spectrum is dominated by the effects brought forth via particle production during the intermediate 50-efoldings of inflation. Thus, if PLANCK does not observe primordial non-Gaussianities, the interaction of inflatons throughout inflation are constrained. If inflatons are identified with moduli fields in string theory, we can thus probe not only the potential along the inflationary trajectory, but also the presence of enhanced symmetries in moduli space in the vicinity of the path the fields take.

\acknowledgments
T.B. and D.B. are grateful for hospitality at the University of Bielefeld and the Institude de AstroParticule et Cosmologie (APC, Paris). CB is grateful for hospitality received from the University of Goettingen. We would like to thank N.~Barnaby for discussions.

\appendix

\section{Conditions for instantaneous particle production \label{app:nonadd}}
We assume particle production to be instantaneous through the paper, but when is this a good approximation? If the non-adiabaticity parameter is large \cite{Kofman:1997yn}
\begin{eqnarray}
\frac{\dot{\omega}}{\omega^2}>1\,,\label{adiabaticity1}
\end{eqnarray}
 where 
 \begin{eqnarray}
\omega(t)\equiv \left( k^2+g^2(\vec{\varphi}-\vec{\varphi}_p)^2\right)^{1/2}\,,\label{adiabaticity}
\end{eqnarray}
explosive particle production similar to pre-heating can take place.
Evaluating the above at $|\vec{\varphi}(t)-\vec{\varphi}_p|=\mu$ and focussing on $k\ll g\mu$ we find the condition
\begin{eqnarray}
\mu\lesssim \sqrt{\frac{v_p}{g}}\equiv \mu_{\mbox{\tiny max}} \label{defmumax}\,,
\end{eqnarray}
where $v_p\equiv|\dot\varphi_p|$. The subscript $p$ denotes the time of particle production when the distance to the ESP is minimal. For larger impact parameters, the approximation in App.~C of \cite{Kofman:2004yc} could be used (see Appendix \ref{app:2}), which provides an iterative procedure under the assumption that 
$\dot{\omega}/\omega^2\ll 1$. However, as we shall see in Sec.~\ref{sec:tunedESP}, the cases where an ESP encounter may be observable, but without observationally ruled out non-Gaussianities, indeed require $\mu\lesssim \mu_{\mbox{\tiny max}}$ anyhow. Then, the production event can be treated as instantaneous, as in Sec.~\ref{sec:grazingESP}. One may further check that the time-scale of particle production is considerably less than the Hubble time, so that the expansion of the universe and the inflationary potential can be neglected during this interval. 

\section{Backreaction \label{app:2}}

The backreaction of  
particle production onto the inflaton is a delicate point. This question is addressed in \cite{Kofman:2004yc} for a two-dimensional scalar field space, without including a potential, and ignoring the cosmological evolution. 

In the present work, we follow the approximation proposed in \cite{Kofman:2004yc}, which  consists of decomposing the impact of particle production into  two successive phases. In the first phase, which is very short, adiabaticity is strongly violated and one can compute, using the methods of preheating,  the number of particles produced, assuming that the trajectory  remains unperturbed. In the second phase, by contrast,  the number of particles is assumed to be conserved (i.e.~simply diluted by the volume expansion in our cosmological case) and the backreaction on the trajectory is taken into account by considering  the interaction between the scalar field and the produced particles, or more specifically the energy density of the particles, given by
\bea
V_{\rm int}=\rho_\chi(\vec\varphi)=g|\vec{\varphi}-\vec{\varphi}_p|n \,,
\eea
where $n$ is the number density of particles and  $m_\chi(\vec\varphi)=g|\vec{\varphi}-\vec{\varphi}_p| $  their effective mass.

For the model discussed in this paper, the dynamical evolution of the inflaton is thus described by the Lagrangian
\beq
{\cal L}=a^3\left[\frac12 \dot\vp^2- V_{\rm int}(|\vec{\varphi}-\vec{\varphi}_p|) -\frac12 m^2\vp^2\right]\, ,
\eeq
with
\beq
V_{\rm int}(|\vec{\varphi}-\vec{\varphi}_p|) =g |\vec{\varphi}-\vec{\varphi}_p| n_p \left(\frac{a_p}{a}\right)^3 \Theta(t-t_p) \,.
\eeq
The associated equation of motion is given by
\beq
\ddot{\vec\varphi}+3H\dot{\vec\varphi}+m^2\vec\varphi=
-\frac{\partial\rho_{\chi}}{\partial\vec\varphi}
=- g n_p\left(\frac{a_p}{a}\right)^3 \frac{\vec\varphi-\vec\varphi_\p}{\|\vec\varphi-\vec\varphi_\p\|}\,.
\eeq

If one decomposes this equation along the directions parallel and orthogonal to the initial (unperturbed) trajectory, one finds
\begin{eqnarray}
\label{eq1}
\ddot{\varphi}_\parallel+3H\dot{\varphi}_\parallel+m^2\varphi_\parallel&=&- g\, n_p\left(\frac{a_p}{a}\right)^3 \Theta(t-t_p) \frac{\varphi_\parallel-\varphi_{\parallel\p}}{\sqrt{(\varphi_\parallel-\varphi_{\parallel\p})^2+(\varphi_\perp-\mu)^2}}\,,
\\
\label{eq2}
\ddot{\varphi}_\perp+3H\dot{\varphi}_\perp+m^2\varphi_\perp&=&- g\,  n_p\left(\frac{a_p}{a}\right)^3 \Theta(t-t_p) \frac{\varphi_\perp-\mu}{\sqrt{(\varphi_\parallel-\varphi_{\parallel\p})^2+(\varphi_\perp-\mu)^2}}\,,
\end{eqnarray}
where we have used the fact that the gradient of the potential vanishes in the orthogonal direction and $\varphi_{\perp\p}=\mu$. (We put the ESP above the $\varphi_\parallel$-axis here, opposite to Fig.~\ref{pic:trajectory1} and Fig.~\ref{pic:trajectory2} but in line with (\ref{eom_perp}) and the numerical plots in this paper).

The numerical results presented in the main body of this paper are based on the above system of equations. Moreover, the approximate equations (\ref{eom0}) and (\ref{eom_perp}),  which  are  
integrated analytically,  are  derived from (\ref{eq1}) and (\ref{eq2}) by assuming  $\varphi_\perp\ll \mu$.

When the above approximation is not valid, in particular when particle production cannot be considered as instantaneous, one must in general solve numerically the  system involving the evolution of the mode functions of the $\chi$ field combined with the equation of motion for the scalar field, as done in \cite{Kofman:2004yc} in the case of a scalar field with $V=0$  in Minkowski spacetime\footnote{Only the regime where the non-adiabaticity  parameter $\dot \omega/\omega^2$ is small seems to be analytically tractable.}.

\end{document}